\documentstyle[12pt]{article}
\makeatletter \oddsidemargin  -.1in \evensidemargin -.1in
\newcommand{\singlespacing}{\let\CS=\@currsize\renewcommand{\baselinestretch}{1}\tiny\CS}
\newcommand{\oneandahalfspacing}{\let\CS=\@currsize\renewcommand{\baselinestretch}{1.25}\tiny\CS}
\newcommand{\doublespacing}{\let\CS=\@currsize\renewcommand{\baselinestretch}{1.35}\tiny\CS}

\newtheorem{rule-def}[theorem]{Rule}

\RequirePackage[dvips]{graphicx} \textheight 22.5cm

\begin{document}

\title {\bf Peristaltic Transport of a Physiological Fluid in an Asymmetric Porous Channel in the Presence of an External Magnetic Field  }
\author{\small J.C.Misra$^1$\thanks{Email address: {\it jcm@maths.iitkgp.ernet.in (J.C.Misra)}}~,~S. Maiti$^2$, G.C.Shit$^1$  \\
\it$^1$Department of Mathematics\\ $^2$\it School of Medical Science and Technology $\&$ Center for Theoretical Studies\\ \it Indian Institute of Technology, Kharagpur-721302, India\\}
\date{}
\maketitle \noindent \doublespacing

\begin{abstract}
The paper deals with a theoretical investigation of the peristaltic transport of a physiological fluid in a porous asymmetric channel under the action of a magnetic field. The stream function, pressure gradient and axial velocity are studied by using appropriate analytical and numerical techniques. Effects of different physical parameters such as permeability, phase difference, wave amplitude and magnetic parameter on the velocity, pumping characteristics, streamline pattern and trapping are investigated with particular emphasis. The computational results are presented in graphical form. The results are found to be in perfect agreement with those of a previous study carried out for a non-porous channel in the absence of a magnetic field.
\end{abstract}

\section{Introduction}
    Peristaltic transport is a natural mechanism of pumping most physiological fluids induced by a progressive wave of area contraction or expansion along the length of the boundary of a fluid-filled distensible tube. Different physiological phenomena, such as the flow of urine from kidney to the bladder through ureters as well as the transport of food material through the digestive tract, semen in the vas differences, fluids in lymphatic vessels, bile from gall bladder into the duodenum, movement of spermatozoa in the ductus efferentes of the male reproductive tract and cervical canal and the transport of ovum in the fallopian tube, the movement of cilia, circulation of blood in small blood vessels, take place by the peristalsis mechanism. This phenomenon is also observed in the propulsion of some industrial fluids.
\begin{center}
\begin{tabular}{|l l|}\hline
{~\bf Nomenclature} &~ \\
~~$a_1,b_1$ & Wave amplitudes\\
~~$B_0 $ & Intensity of the magnetic field\\
~~$d_1,d_2 $ & Width of the channel \\
~~$H_1,H_2 $ & Vertical displacements of the upper and lower walls respectively\\
~~$M $ & Magnetic parameter\\
~~$P $ & Fluid pressure\\
~~$Q $ & Flux at any axial station\\
~~$R $ & Reynolds number of the fluid\\
~~$t $  & Time\\
~~$U,V$ & Velocity components in X,Y directions respectively\\
~~$X,Y $ & Rectangular Cartesian co-ordinates\\
~~$\alpha $ & Function defined in equation (12)\\
~~$\delta $ & Wave number\\
~~$\Delta p$ & Pressure difference between the channel ends\\
~~$\lambda $ & Wave length of the travelling wave motion of the wall\\
~~$\mu $ & Viscosity of the fluid\\
~~$\nu $ & Kinematic viscosity of the fluid\\
~~$\phi$ & Phase difference\\
~~$\rho$ & Density of the fluid \\
~~$\psi$ & Stream function\\
~~$\sigma $ & Electrical conductivity of the fluid\\
\hline
\end{tabular}
\end{center}

 This mechanism finds applications in roller and finger pumps, heart-lung machine, blood pump machine, dialysis machine and also noxious fluid transport in nuclear industries. Some of the recent important theoretical studies on peristaltic transport include those by Usha and Rao \cite{r1}, Misra and Pandey \cite{r2,r3,r4,r5,r6,r7}, Mishra and Rao \cite{r8, r9}, Mekheimer \cite{r10}, Afifi \cite{r11}, Eytan et al. \cite{r12}, Eldabe et al. \cite{r13}, Hakeem et al. \cite{r14} as well as Hayat and Ali \cite{r15}. These studies have been carried out either by using lubrication theory neglecting the fluid inertia and wall curvature without any limitation of wave amplitude or by considering small peristaltic wave amplitude with arbitrary Reynolds number. Using the assumptions of thin shell and lubrication theories, Carew and Pedley \cite{r16} put forward a mathematical analysis for the study of peristaltic pumping in the ureter. Antanovskii and Ramkissoon \cite{r17} also took the help of lubrication theory in order to study the peristaltic transport of a compressible viscous fluid through a pipe when the pressure drop changes with time, by taking into account the wall deformation of the pipe. Lee and Fung \cite{r18} studied theoretically the peristaltic phenomenon observed in blood flow, while Lee  \cite{r19,r20} reported his experimental studies on peristalsis observed in large intestines. \\
          Taylor  \cite{r21} conducted a theoretical study on asymmetric wave propagation in waving sheets with an object to put forward an explanation to the mechanical interaction between neighbouring spermatozoa. By considering a biharmonic stream function in the form of a Fourier series and retaining only the starting term of the stream function series, he made an attempt to simulate cases for symmetric /asymmetric wave propagation. The boundary integral method along with a suitable numerical technique was employed by Pozrikidis \cite{r22} to study the fluid dynamics involved in the problem of peristaltic transport in an asymmetric channel under Stokes flow conditions. A mathematical analysis for peristaltic transport of an incompressible Newtonian fluid was carried out by Mishra and Rao \cite{r8} as well as  Elshehaway et al.\cite{r23}. In both these studies, channel flow was considered where the channel is asymmetric. These two studies are similar in nature, however unlike \cite{r8}, the channel was considered to be porous in \cite{r23}. The work reported in \cite{r8} was extended by Reddy et al. \cite{r24} to the propagation of peristaltic transport of a power-law fluid in an asymmetric channel. This theoretical study was based on a similar study conducted earlier by Reddy et al. \cite{r25} for the propagation of symmetric peristaltic waves in a Newtonian fluid model. \\
          Recent physiological researches (Cf. Vries et al.\cite{r26}) indicate that uterine peristalsis resulting from myometrial contraction can take place in both symmetric and asymmetric directions. In different studies pertaining to the gastrointestinal tract, intra-pleural membranes, capillary walls, human lung, bile duct, gall bladder with stones and small blood vessels, flow in porous tubes and deformable porous layers have been examined by various authors. It is known that the primary function of gastrointestinal tract consists of absorbing nutrients from the mix of food and liquid that passes through the tract. As mentioned by Keener and Sneyd \cite{r27}, gastrointestinal tract is surrounded by a number of heavily innervated muscle layers which are smooth muscles consisting of many folds. There exist pores in the junctions between them, although the junctions are tight. Also, Bergel \cite{r28} observed that the capillary walls are surrounded by flattened endothelial cell layers which are porous. \\
          Currently studies on peristaltic motion in magneto hydrodynamic (MHD) flows of electrically conducting physiological fluids have become a subject of growing interest for researchers. This is due to the fact that such studies are useful particularly for having a proper understanding of the functioning of different machines used by clinicians for pumping blood and magnetic resonance imaging (MRI). The related studies also contribute to updating the information that is required to operate machines like MHD peristaltic compressors. As pointed out by Agrawal and Anwaruddin \cite{r29}, theoretical researches with an aim to explore the effect of a magnetic field on the flow of blood in atherosclerotic vessels  also find application in a blood pump used by cardiac surgeons during the surgical procedure. Investigations related to the study of the effect of a magnetic field on the movement of blood indicate that if we apply a frequency pulsing magnetic field of low intensity in a controlled manner, it is possible to modify the behaviour of cells and tissues. Li et al. \cite{r30} observed that an impulsive magnetic field can be used as a theraptic means to treat patients who have stone fragments in their urinary tract. \\
          In view of all the aforesaid observations and remarks, the present theoretical study has been designed in such a manner that it can explore a variety of information that can enrich the information bank required for magneto therapy, which consists of applying a magnetic field for the treatment of various diseases. From the physiological fluid dynamics point of view, it is a study of the peristaltic transport of a physiological fluid in a porous asymmetric channel under the action of an external magnetic field. The peristaltic wave train on the channel wall has been considered to have different amplitudes and phases in view of the variation in the channel width, wave amplitude and phase difference. It is believed that the results presented here will find prospective application in the study of various fluid mechanical problems associated with gastrointestinal tract, intra pleural membranes, capillary walls and small blood vessels.  

\section{The Problem and its Solution}
    Let us consider the  peristaltic motion of an incompressible viscous fluid through a porous medium channel under the action of a magnetic field. Let $ Y=H_1$ and $Y=H_2$ be respectively the upper and lower boundaries of the channel. The medium is considered to be induced by a sinusoidal wave train propagating with a constant speed $c$ along the channel wall, such that \\
$Y=H_1=d_1+a_1 cos(\frac{2\pi}{\lambda}(X-ct)) $ on the upper boundary \\
 and \\$Y=H_2=-d_2-b_1 cos(\frac{2\pi}{\lambda}(X-ct)+\phi) $ on the lower boundary,\\
where $a_1,~b_1$ are the wave amplitudes , $\lambda$ is the wave length, $d_1+d_2$ the width of the channel, $\phi$ the phase difference ($0\le \phi \le \pi $) of the waves. $\phi=0$ corresponds to the symmetric channel with waves out of phase and $\phi =\pi$ represents the waves in phase; quantities $a_1,~b_1,~d_1,~d_2,$ and $\phi$ are  such that \\
\begin{equation}
a_1^2+b_1^2+2a_1b_1 cos(\phi) \le (d_1+d_2)^2 
\end{equation} \\
The equations that govern the fluid motion in a porous medium under the influence of a magnetic field are: 
\begin{equation}
\frac{\partial U}{\partial X}+\frac{\partial V}{\partial Y}=0
\end{equation}
\begin{equation}
\frac{\partial U}{\partial t}+U\frac{\partial U}{\partial X}+V\frac{\partial U}{\partial Y}=-\frac{1}{\rho}\frac{\partial P}{\partial X}+\nu \left (\frac{\partial^2 U}{\partial X^2}+\frac{\partial^2 U}{\partial Y^2}\right )-\frac{\nu_1}{k}U-\frac{\sigma B_0^2}{\rho}U
\end{equation}
\begin{equation}
\frac{\partial V}{\partial t}+U\frac{\partial V}{\partial X}+V\frac{\partial V}{\partial Y}=-\frac{1}{\rho}\frac{\partial P}{\partial Y}+\nu \left (\frac{\partial^2 V}{\partial X^2}+\frac{\partial^2 V}{\partial Y^2}\right )-\frac{\nu_1}{k}V
\end{equation}
where $\nu_1=\frac{\nu}{\beta}$, k is the permeability parameter, $B_0$ the intensity of the magnetic field acting along the y-axis and the induced magnetic field is assumed to be negligible. \\
Considering a wave frame $(x,y)$ moving with a velocity $c$ away from a fixed frame $(X,Y)$, let us the transformations \\
\begin{equation}
x=X-ct, ~~y=Y, ~~ u=U-c, ~~ v=V, ~~ p(x)=P(X,t)
\end{equation}
in which  $(u,v)$ and $(U,V)$ are the velocity components, $p$ and $P$ stand for  pressure in wave and fixed frame of reference respectively.\\
If $\psi$ be the stream function, we have\\ $u=\frac{\partial \psi}{
\partial y}, ~~ v=-\frac{\partial \psi}{\partial x} $.\\
 Now eliminating P between  (3) and (4), we obtain the governing equation expressed in terms of $\psi$. Thus the equation governing the steady flow of the fluid in the wave frame of reference can be written in the form 
 \\ \begin{eqnarray}
 \psi_y\psi_{yyx}-\psi_x\psi_{yyy}+ \psi_y\psi_{xxx}-
\psi_x\psi_{xxy} &=& \nu (\psi_{yyyy}+2\psi_{xxyy}+\psi_{xxxx}) \nonumber \\  &-&\frac{\nu}{\beta k}\psi_{yy}-\frac{\nu}{\beta k}\psi_{xx} -\frac{\sigma B_0^2}{\rho}\psi_{yy}
\end{eqnarray}
  In the sequal, we shall make use of the following non-dimensional variables: 
\begin{eqnarray}
&\bar{x}&=\frac{x}{\lambda}, ~~\bar{y}=\frac{y}{d_1}, ~~\bar{u}=\frac{u}{c},
~~\bar{v}=\frac{v}{c\delta}, ~~ \delta=\frac{d_1}{\lambda}, ~~\bar{p}=\frac{d_1^2p}{\mu c\lambda},~~
\bar{t}=\frac{ct}{\lambda}, ~~h_1=\frac{H_1}{d_1}, ~~h_2=\frac{H_2}{d_1}, ~~d=\frac{d_2}{d_1},\nonumber \\
&a&=\frac{a_1}{d_1}, ~~b=\frac{b_1}{d_1}, ~~ R=\frac{c d_1}{\nu }, ~~ \bar{\psi}=\frac{\psi}{cd_1}, ~~
\bar{k}=\frac{k}{d_1^2}, ~~ M=\frac{\sigma }{\mu}B_0^2d_1^2(>2)
\end{eqnarray}
  Henceforward we shall use the non-dimensional variables defined in (7). If we now drop bars over the symbols, the equation governing the flow of the fluid reads\\ \begin{eqnarray}
R\delta \left \{\psi_y\psi_{yyx}-\psi_x\psi_{yyy}+\delta^2 \left (\psi_y\psi_{xxx}-
\psi_x\psi_{xxy} \right ) \right \}&=&\psi_{yyyy}+2\delta^2 \psi_{xxyy}+\delta^4 \psi_{xxxx} \nonumber \\
&-&\frac{1}{\beta k}\psi_{yy}-\frac{\delta^2}{\beta k}\psi_{xx}-M\psi_{yy}
\end{eqnarray}

{\bf Boundary Conditions:}\\
  Considering the physics of the problem, the boundary conditions may be put mathematically as \\
\begin{eqnarray}
(i)~~~~~~~ \psi=\frac{q}{2}~~ at~~ y=h_1=1+a~ cos(2\pi x )~~~~~~~~~~~~~~~~~~~~ \nonumber \\
    \psi=-\frac{q}{2} ~~ at ~~ y=h_2=-d-b~cos(2\pi x+\phi) \nonumber~~~~~~~~~~~ \\
(ii) ~~~~~\frac{\partial \psi}{\partial y}=-1 ~~ at~~~ h_1,~h_2~~~~~~~~~~~~~~~~~~~~~~~~~~~~~~~~~~~~~~~~
\end{eqnarray}
where $q$ refers to the flux in the wave frame and $a,~b,~d,~\phi $ are such that \\
\begin{equation}
a^2+b^2+2abcos(\phi) \le (1+d)^2
\end{equation}
           Applying long wave length approximation $(\delta \ll 1 )$ and assuming the Reynolds number to small, the equation (8) approximates to 
\begin{eqnarray}
\psi_{yyyy}-\alpha^2\psi_{yy}=0
\end{eqnarray}
where  \begin{eqnarray}\alpha^2=\frac{1}{\beta k}+M\end{eqnarray}
          The solution of equation (11) satisfying (9) is
\begin{equation}
\psi =C_1+C_2y+C_3e^{\alpha y}+C_4e^{-\alpha y},
\end{equation}
with 
\[C_1=\frac{q}{2}+h_1-\frac{\alpha h_1(q+h_1-h_2)}{\alpha (h_1-h_2)-2tanh\frac{\alpha}{2}(h_1-h_2)}+
\frac{(q+h_1-h_2)(e^{\alpha (h_1-h_2)}-1)(1-e^{-\alpha (h_1-h_2)})}{4(1-cosh\alpha(h_1-h_2)
)+2\alpha (h_1-h_2)sinh \alpha (h_1-h_2)} \]
\[C_2=-1+\frac{\alpha sinh \alpha(h_1-h_2)(q+h_1-h_2)}{2(1-cosh\alpha(h_1-h_2)
)+\alpha (h_1-h_2)sinh \alpha (h_1-h_2)} \] 
\[C_3=-\frac{(q+h_1-h_2)e^{-\alpha h_1}(e^{\alpha (h_1-h_2)}-1)}{4(1-cosh\alpha (h_1-h_2)
)+2\alpha (h_1-h_2)sinh \alpha (h_1-h_2)} \]
\[C_4=\frac{(q+h_1-h_2)e^{\alpha h_2}(e^{\alpha (h_1-h_2)}-1)}{4(1-cosh\alpha (h_1-h_2)
)+2\alpha (h_1-h_2)sinh \alpha (h_1-h_2)} \]
where $h_2 \le y \le h_1 $. \\
           In the absence of the magnetic field, these results are found to coincide with those obtained in  \cite{r8}, where the porosity effect is disregarded.\\
The flux at any axial station in the fixed frame is found to be given by
\[ Q=\int_{h_2}^{h_1}(u+1)dy =q+h_1-h_2, \]
        while the expression for the volumetric flow rate over one period $T(=\frac{\lambda }{c} $) of the peristaltic wave is obtained as 
\[ \bar{Q}=\frac{1}{T}\int_{0}^{T}Qdt=\frac{1}{T}\int_{0}^{T}(q+h_1-h_2)dt=q+1+d \]
          The pressure gradient obtained from the dimensionless momentum equation for axial velocity can be expressed as  
\[ \frac{dp}{dx}=\psi_{yyy}-\alpha^2 \psi_y-\alpha^2 \]
\[=-\frac{\alpha^3sinh \alpha(h_1-h_2)(q+h_1-h_2)}{2(1-cosh\alpha(h_1-h_2)
)+\alpha (h_1-h_2)sinh \alpha (h_1-h_2)} \] 
        On integration over one wave length it reduces to 
\[\Delta p=-\alpha^3 \bar{Q}I_1-\alpha^3I_2 \]
where 
\[ I_1=\int_{0}^{1}\frac{sinh \alpha(h_1-h_2)dx}{2(1-cosh\alpha(h_1-h_2)
)+\alpha (h_1-h_2)sinh \alpha (h_1-h_2)} \]  
\[ I_2=\int_{0}^{1}\frac{sinh \alpha(h_1-h_2)(h_1-h_2-1-d)dx}{2(1-cosh\alpha(h_1-h_2)
)+\alpha (h_1-h_2)sinh \alpha (h_1-h_2)} \]  
       In the fixed frame of reference, the expression for the axial velocity is
\[U(X,Y,t)=\frac{\alpha(q+h_1-h_2)(cosh\frac{\alpha}{2}(h_1-h_2)-cosh\frac{\alpha}{2}(h_1+h_2-2Y))}{\alpha(h_1-h_2)cosh\frac{\alpha}{2}(h_1-h_2)-2sinh\frac{\alpha}{2}(h_1-h_2)}~\]\\
with 
\[h_1=1+a~cos[2\pi(X-t)]~~ and~~ h_2=-d-b~cos[2\pi(X-t+\phi)].\]

\section{Results and Discussions}
     Analytical expressions for the pressure gradient, the volumetric flow rate, axial velocity and the stream function for the problem under consideration have been calculated and presented in the previous section. In this section, the problem has been investigated further by employing appropriate computational method and numerical techniques. For the purpose of computation, the following set of values (which are in agreement with experimental data for physiological fluids) of the different parameters involved in the forgoing theoretical analysis have been used: \\
a, b= 0.3, 0.4, 0.5, 0.7, 1.2; d=1.0, 2.0; $\bar{Q} =1.4$; M=0 to 8; $\beta$k=0.02, 0.05, 0.08, 0.1, 0.13, 0.15, 0.2, 0.5, 1, 10, 100, 1000; 
$\phi =0.0, \frac{\pi}{2},  \frac{\pi}{4},  \frac{5\pi}{6},  \pi$. Small values of the permeability factor refer to cases when the medium is almost rigid.\\
              Due to the complexity of the problem under consideration, it was not possible to determine the analytical expression of $\Delta p$ in terms of  $\bar{Q}$ and other parameters. Thus had to resort to the use of appropriate numerical method along with quadrature formula. Similarly, an analytical treatment was found inadequate for finding the velocity as a function of $\Delta p$, X, Y, t and other parameters that are involved. This necessitated calculating first\\
\[\bar{Q}=-\frac{\Delta p +\alpha^3I_2}{\alpha^3I_1},\]\\
which was also a very complicated task. Here too, we had to make use of numerical quadrature.\\

{\bf Velocity Change}\\
 The distribution of axial velocity in the case of free pumping for different values of the phase difference, the permeability parameter k and the magnetic parameter M is shown in Figs 2-5. Fig.2 shows that the results computed on the basis of our study in absence of any external magnetic field (M=0) perfectly match with the results reported in \cite{r8}. Figs.3(a)-(c) give the changes in the velocity distribution when M=2, as the phase difference  $\phi $ changes from 0 to  $\pi$. These figures indicate that the velocity changes considerably with change in the values of the amplitudes of the peristaltic wave at the upper and lower walls of the channel. These figures reveal that the magnitude of the velocity gradually decreases and the extent of flow reversal near the boundary increases as the phase difference increases, when the wave amplitude at the upper boundary of the channel is more than or equal to the wave amplitude at the lower boundary. It is interesting to note that when the wave amplitude at the lower boundary exceeds that at the upper boundary, the scenario is similar when  $\phi $ changes from 0 to $\frac{\pi}{2}$. In this case, when  $\phi $ exceeds $\frac{\pi}{2}$, the flow reversal at the boundary starts decreasing and reduces to zero when $\frac{3\pi}{4}$. When $\phi $ increases further, the flow reversal takes place in the central region and the velocity near the boundary increases.\\         
          Fig. 4 shows that flow reversal takes places for all the values of the magnetic parameter examined here and also that while in the central portion of the channel, the axial velocity decreases with the increase of magnetic intensity, the behaviour is slightly opposite in the lower and upper regions. It may be noted from Fig. 5 that in central region of the channel the velocity increases with the increase in the permeability parameter of the channel, while an opposite trend is found in the lower and upper regions. The occurrence of reverse flow at the initial stage is found to be independent of the permeability factor.\\

{\bf Pumping Characteristics}\\
 Figs. 6-8 illustrate the variation of the volumetric flow rate of peristaltic wave with pressure gradient for different values of the phase difference, permeability factor and the magnetic intensity. One may observe from Fig. 6 that in the range of values of the pressure gradient examined in the present study, in the entire pumping region ($\Delta p >$0), in the free pumping zone ($\Delta p =$0) as well as in the co-pumping region ($\Delta p <$0) for $\Delta p >$-12.3, the volumetric flow rate decreases with the increase in the phase difference. However, the trend reverses as soon as the pressure gradient drops below -12.3 . These observations, of course, hold good for a particular set of values indicated in  Fig. 6 caption. Fig. 7 shows that in the entire pumping region the volumetric flow rate decreases with the increase in permeability parameter, where as in the co-pumping region, a reverse trend is noticed. One can also observed that the volumetric flow rate can be gradually increased in the pumping region and gradually reduced in the co-pumping region by increasing the intensity of the applied magnetic field.\\

{\bf Trapping}\\
    A theoretical study on peristaltic pumping of a fluid was conducted by Shapiro et al. \cite{r31} for the case when the wave length is large and the Reynolds number is small. This study revealed that at high flow rates and large occlusions, there is a region of closed stream lines in the wave frame, whereby some fluid gets trapped within a wave of propagation. The trapped fluid mass is found to move with a mean speed equal to that of the wave. Thus the phenomenon of trapping may be looked upon as the formation of an internally circulating bolus of fluid. Owing to the trapping phenomenon, there will exist stagnation points, where both the velocity components of the fluid vanish in the wave frame. The stagnation points will be located at the points where the central line of the channel intersects the curve $\psi=0$ ($\psi$ being the stream function). In case of asymmetric channel flow, however, some stagnation points may be located elsewhere.\\ Figs. 9(a)-(d) reveals a very interesting fact that with the same amplitude, a=b=0.5 for $\bar{Q} =1.4$, the bolus appearing in the central region for $\phi =0$ moves towards left and decreases in size as $\phi $ increases. For $\phi= \pi$, the bolus completely disappears and the stream lines are parallel to one another.  \\
           The influence of the intensity of the applied magnetic field on the streamlines for peristaltic transport velocity where a=b=0.5 and $\beta$k=1.0, $\phi $=0.0 and d=1.0 is demonstrated in Figs. 10(a)-(d). These figures indicate that when the magnetic field intensity is low, bolus is formed at the central region of the channel. These bolus is shifted towards the upper boundary of the channel with the increase in the magnetic field strength. Fig. 10(d)  reveals that the bolus completely vanishes when M $\ge $7.0. \\
           Figs. 11(a)-(d) indicate that when M=1, in the absence of phase difference, bolus is formed for greater values of the permeability factor but no bolus is formed when the permeability parameter of the channel is small.\\ 
            A comparison between Fig. 11(d) and Fig. 12(a) reveals that when M=1, bolus formation is of different type when the phase difference is   $\frac{\pi}{4}$, even the channel has the same permeability. But when the intensity of the magnetic field is higher, the bolus disappears even if the phase difference and permeability remain unchanged [cf. Fig. 12(b)]. Figs. 13(a)-(b) give us an idea of the type of bolus formation for greater values of the permeability factor. The variation in the shape of the streamlines and bolus formation for different sets of values of a and b can been seen from Figs. 14(a)-(c).

\section{Summary and Conclusion}
 The peristaltic transport of a Newtonian fluid (as an idealization of a physiological fluid) in a porous asymmetric channel under the action of an externally applied magnetic field has been the concern  of the present investigation. The study particularly pertains to the situation when the Reynolds number is low and curvature of the channel is quite small. Emphasis has been  paid to study some important bio-fluid dynamical phenomena of peristaltic transport  such as the pumping characteristics, axial velocity as well as streamline patterns and trapping as a function of asymmetric wave motility parameters, the permeability parameter and the magnetic parameter. All these have been treated both analytically and numerically. The results and observations are good agreement with those reported in \cite{r8, r9, r10, r32}.\\
             On the basis of the observations of the present study, one can make an important conclusion that it is possible to increase pumping for the adverse pressure gradient by applying an external magnetic field and that the bolus formation  can be avoided by suitably adjusting the magnetic field intensity. The present study reveals further that the velocity change is strongly dependent on the phase difference, amplitudes of  wave propagating in the upper and lower regions, magnetic field intensity and the channel permeability.\\  
           The results throw some light on problems associated with fluid movement in the gastrointestinal tract, intra-uterine fluid motion induced by uterine contraction as well as flow through small blood vessels and intra-pleural membranes. \\
  The study suggest that a convenient criterion for the presence of a trapped bolus is the existence of stagnation points in the wave frame. This physical phenomenon can be considered responsible for formation of thrombus in blood and the movement of food bolus in the gastrointestinal tract.\\

{\bf Acknowledgment:} {\it The authors wish to express their deep sense of gratitude to the anonymous referees for their appreciation of the piece of research embodied in the paper, for their words of praise regarding the results reported in the paper and also for their suggestions based upon which the revised manuscript has been prepared. The authors are also thankful to The Council of Scientific and Industrial Research (CSIR), New Delhi for their financial support towards this study.}

\newpage
\begin{minipage}{1.0\textwidth}
     \begin{center}
       \includegraphics[width=5.15in,height=3.2in]{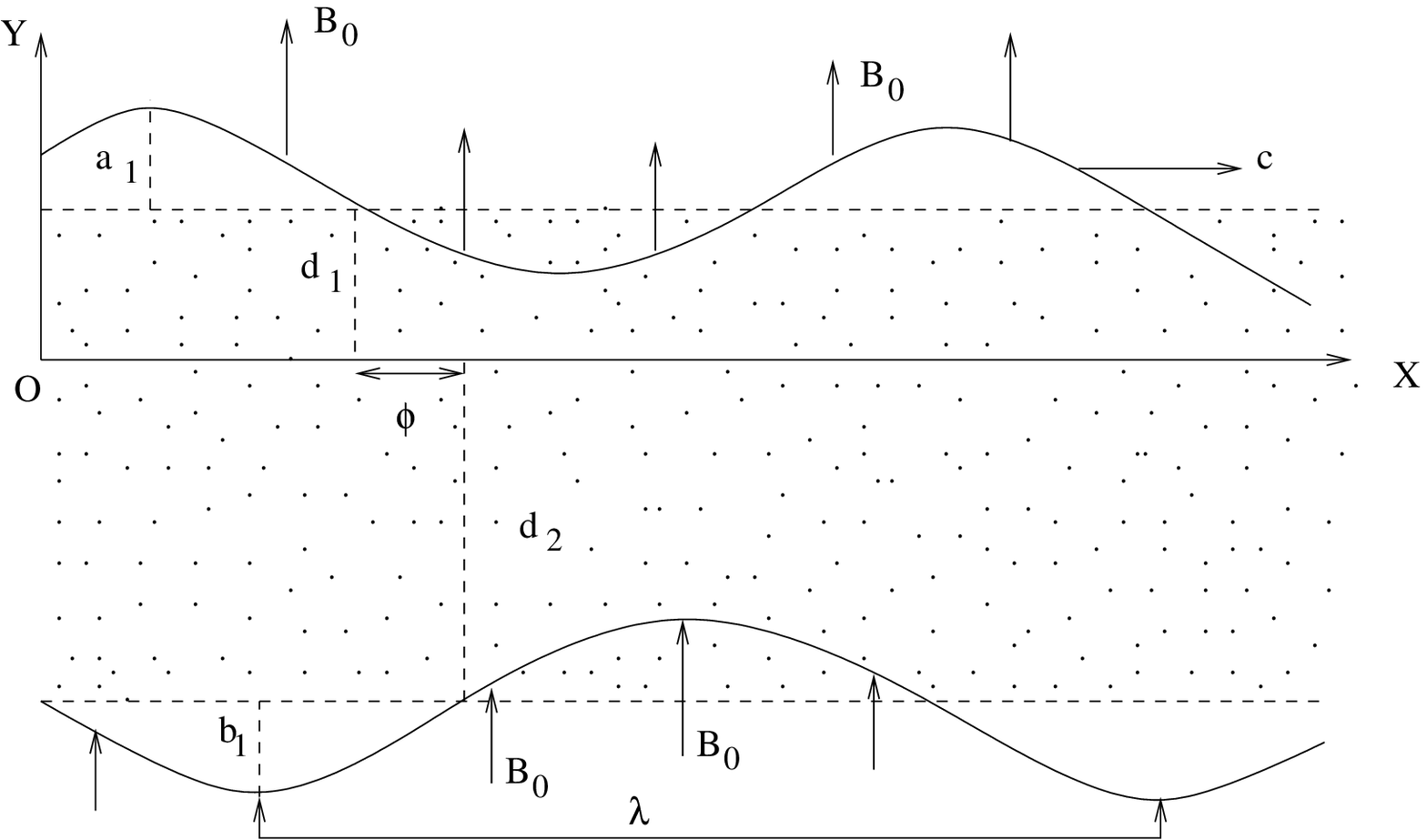} \\
 Fig. 1: A physical sketch of the problem \\
 \includegraphics[width=5.15in,height=3.2in]{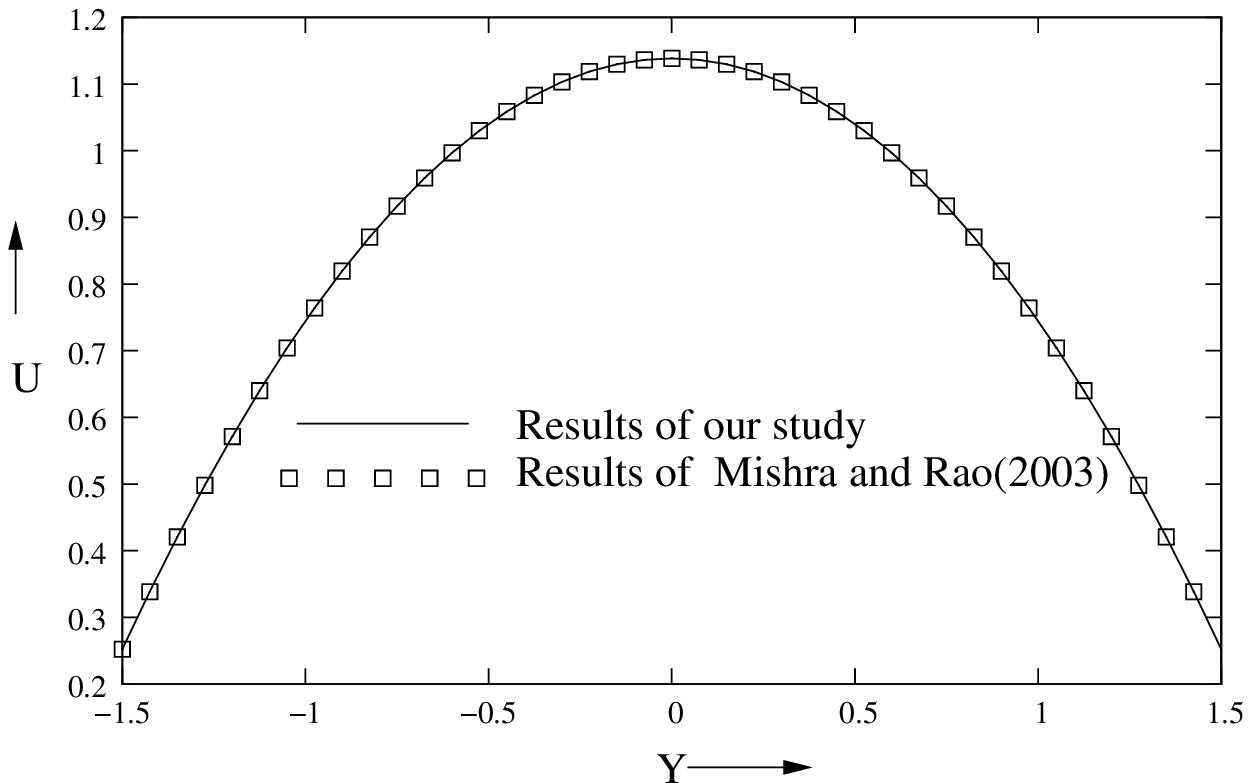} \\
 Fig. 2: Distribution of axial velocity for X=1, t=1, a=0.7, b=0.7, d=1, $\phi$=0, $\beta$k=1000, $\Delta p$=0, M=0   \\
\end{center}
\end{minipage}\vspace*{.25cm}
\newpage

\begin{minipage}{1.0\textwidth}
     \begin{center}
       \includegraphics[width=4.8in,height=3.2in]{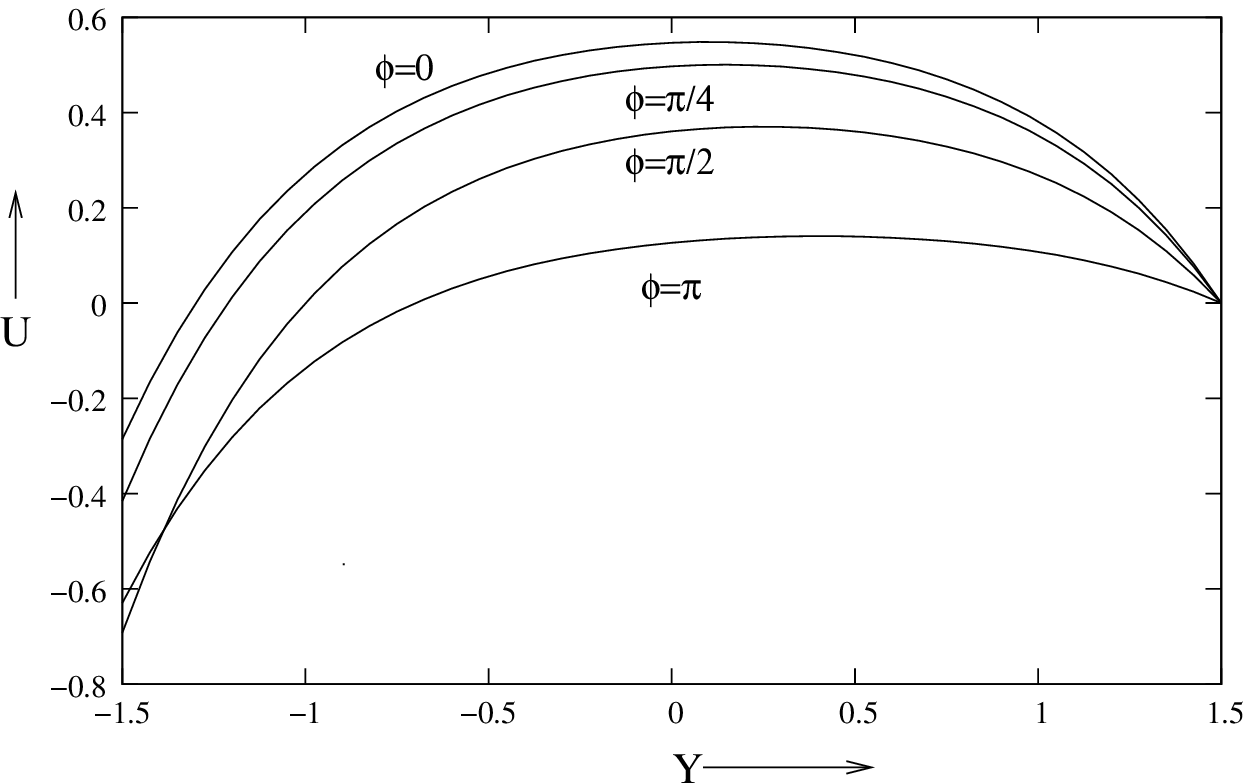} \\
 Fig. 3a: Distribution of axial velocity for different values of $\phi$, where X=1, t=1, $\Delta p$=0, a=0.5, b=0.3, d=1, $\beta$k=0.8, M=2 \\
\includegraphics[width=4.8in,height=3.2in]{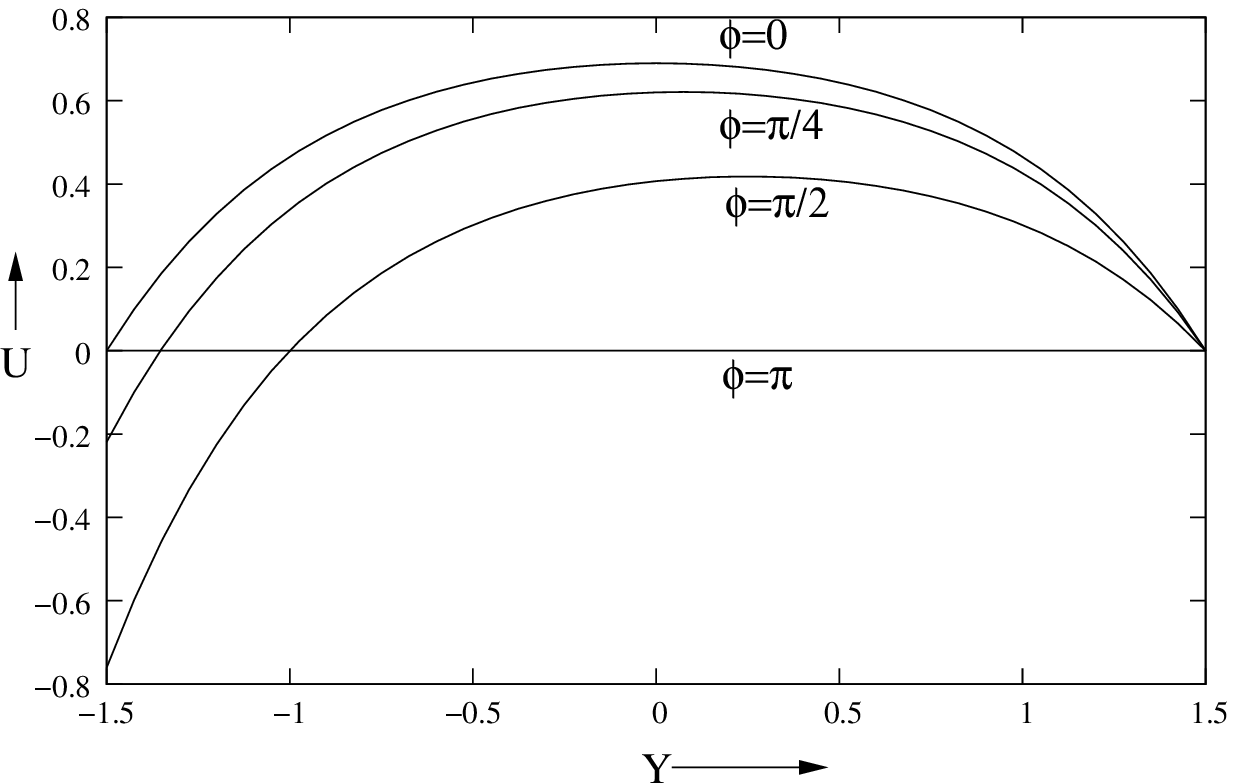} \\
 Fig. 3b: Distribution of axial velocity for different values of $\phi$, where X=1, t=1, $\Delta p$=0, a=0.5, b=0.5, d=1, $\beta$k=0.8, M=2 \\
\end{center}
\end{minipage}\vspace*{.25cm}
\newpage
\begin{minipage}{1.0\textwidth}
     \begin{center}
       \includegraphics[width=4.8in,height=3.2in]{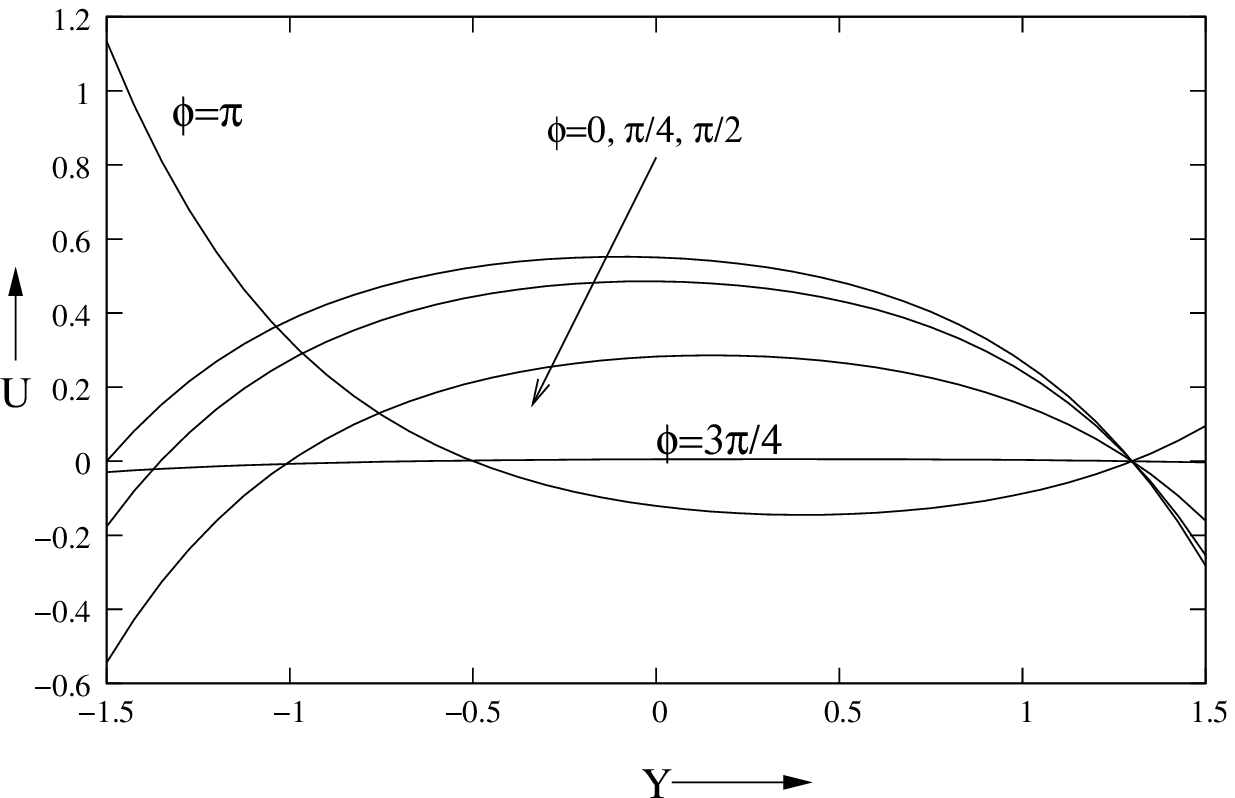} \\
 Fig. 3c: Distribution of axial velocity for different values of $\phi$, where X=1, t=1, $\Delta p$=0, a=0.3, b=0.5, d=1, $\beta$k=0.8, M=2 \\
\includegraphics[width=4.8in,height=3.2in]{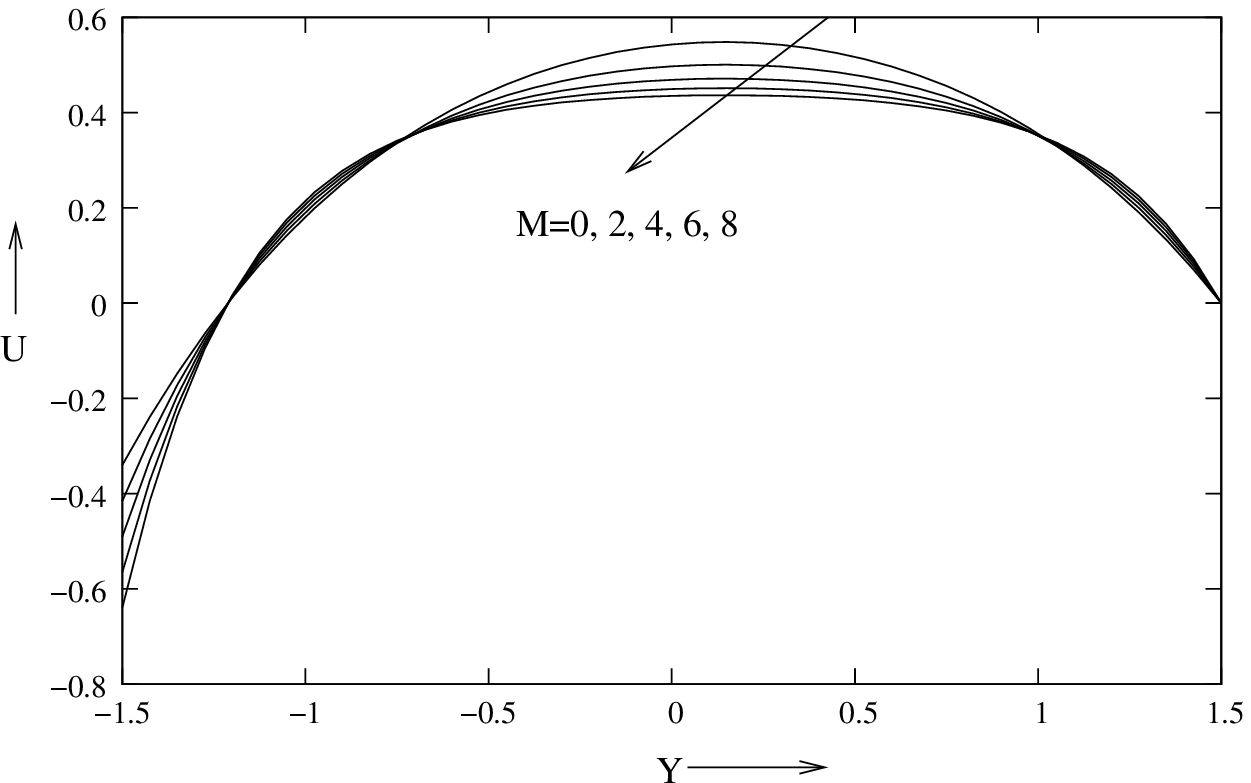} \\
  Fig. 4: Distribution of axial velocity for different  values of M, where X=1, t=1, $\Delta p$=0, a=0.5, b=0.3, d=1, $\beta$k=0.8, $\phi$=$\frac{\pi}{4}$ \\
\end{center}
\end{minipage}\vspace*{.25cm}
\newpage
\begin{minipage}{1.0\textwidth}
     \begin{center}
       \includegraphics[width=4.8in,height=3.2in]{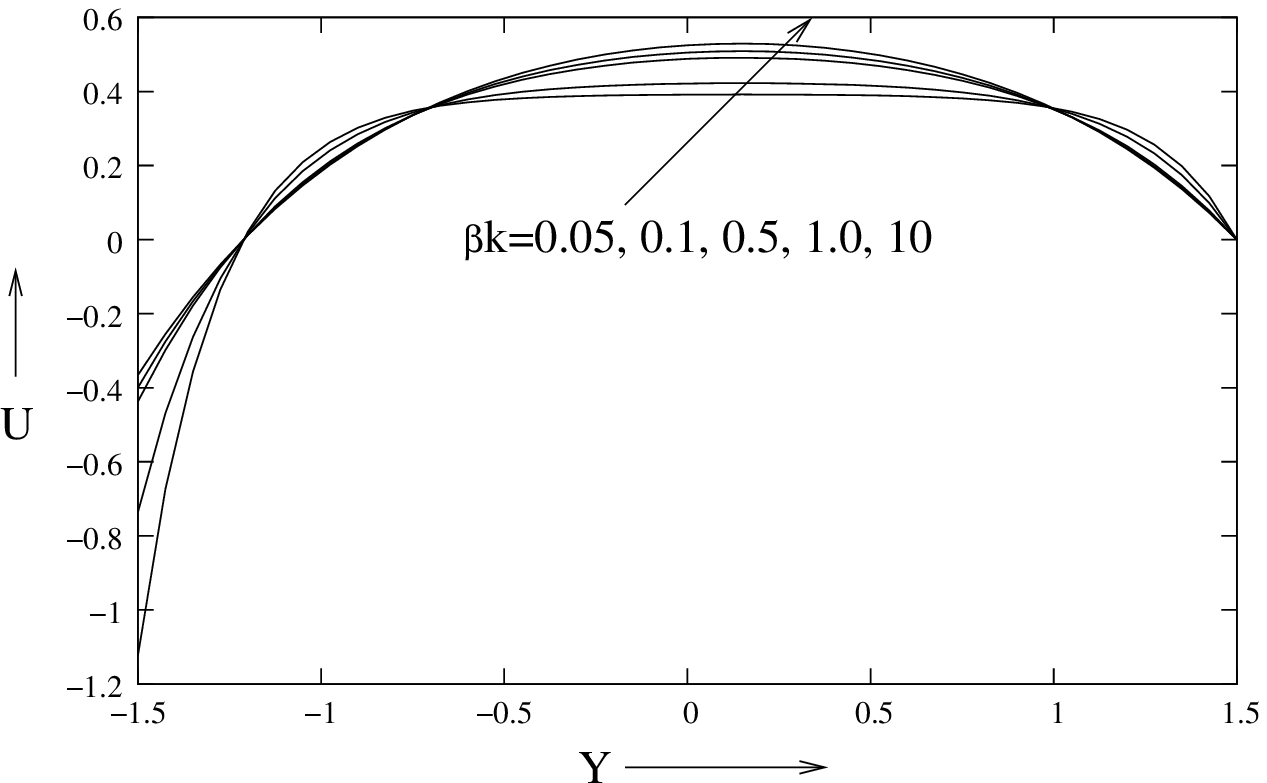} \\
 Fig. 5:  Distribution of axial velocity for different values of $\beta$k, where X=1, t=1, $\Delta p$=0, a=0.5, b=0.3, d=1, M=2, $\phi$=$\frac{\pi}{4}$ \\
\includegraphics[width=4.8in,height=3.2in]{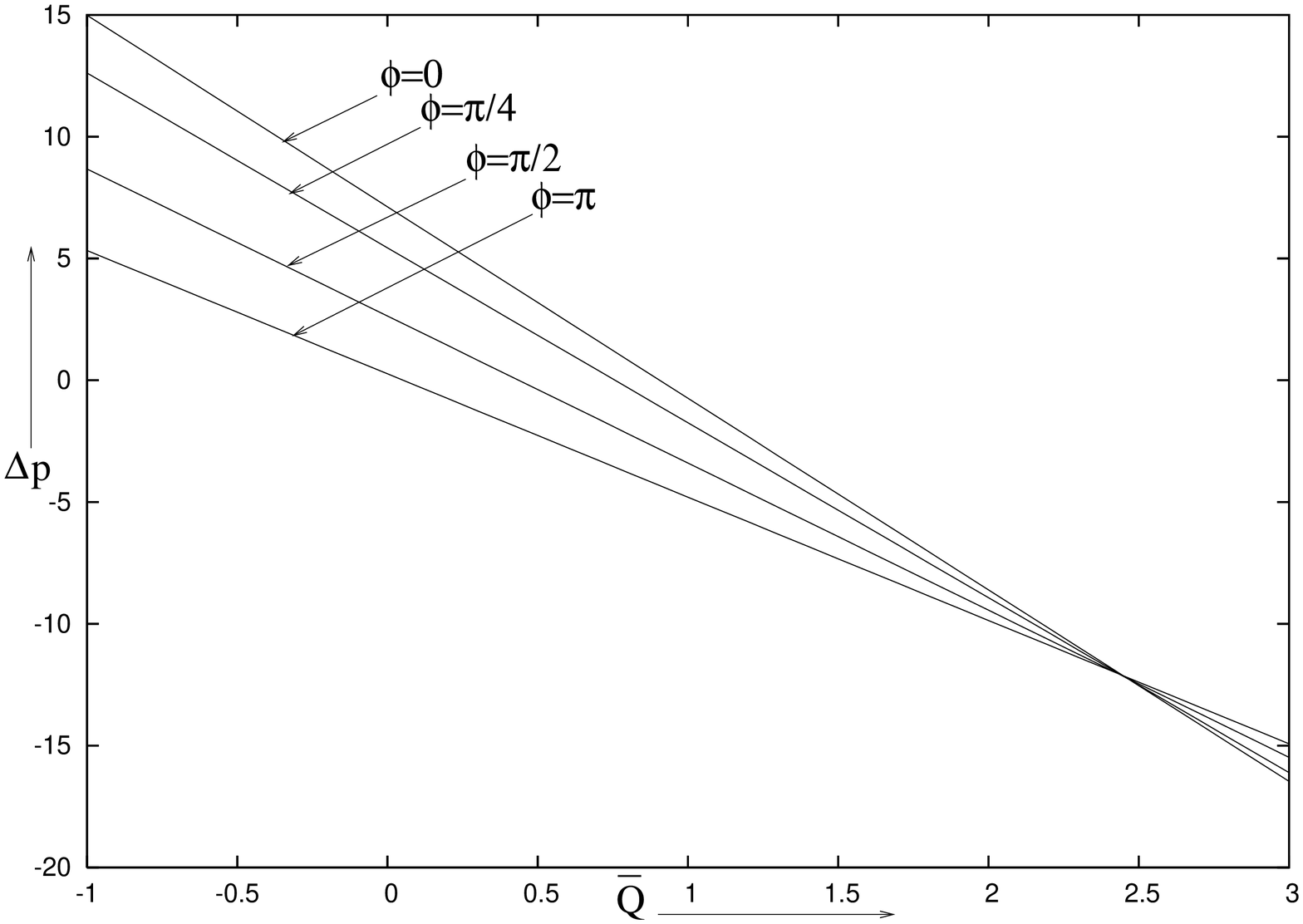} \\
 Fig. 6: Variation of $\bar{Q}$ with  $\Delta p$ for different values of $\phi$ along with a=0.7, b=1.2, d=2, $\beta$k=0.1, M=2 \\
\end{center}
\end{minipage}\vspace*{.25cm}
\newpage
\begin{minipage}{1.0\textwidth}
     \begin{center}
       \includegraphics[width=4.8in,height=3.2in]{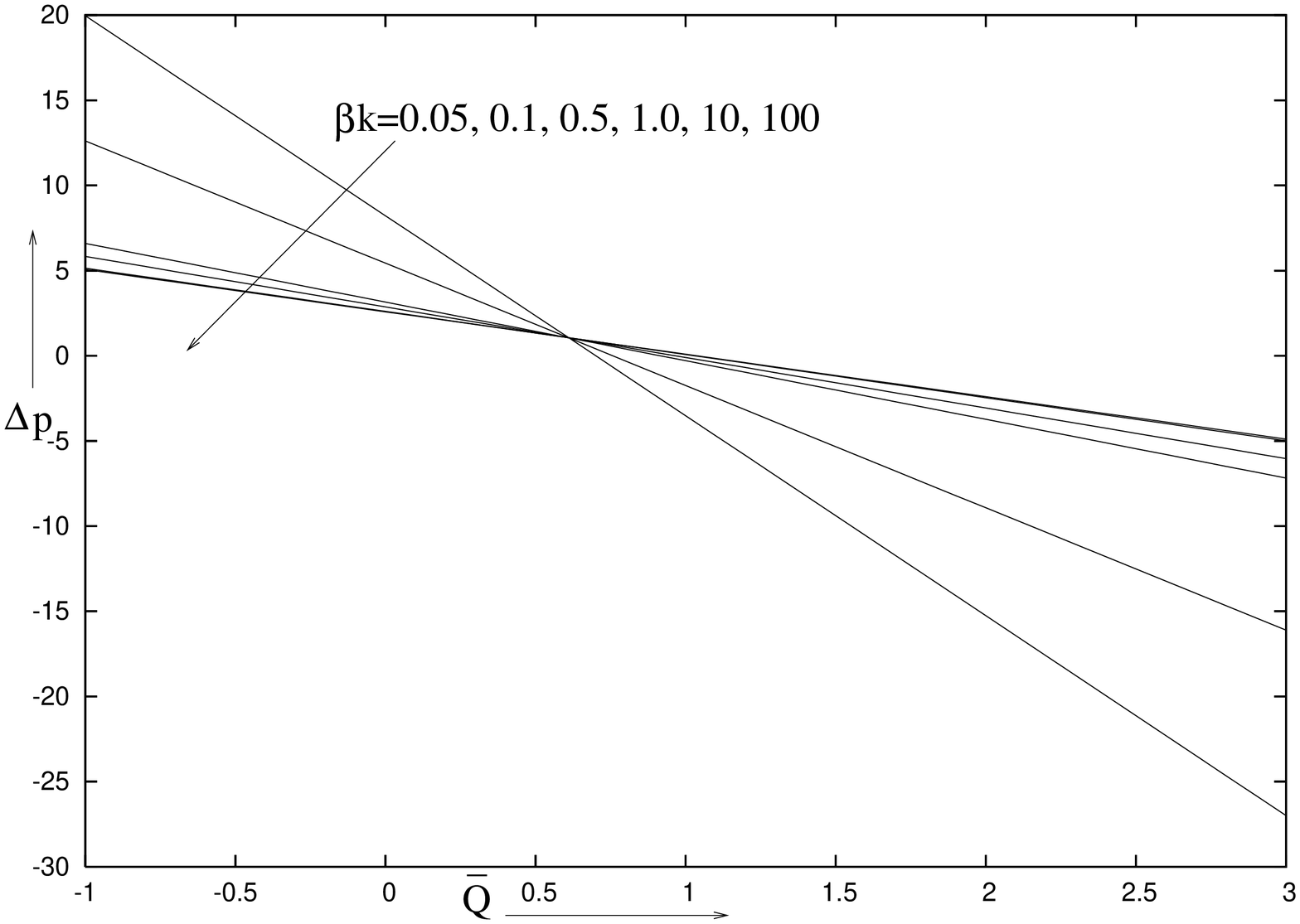} \\
 Fig. 7: Variation of $\bar{Q}$ with  $\Delta p$ for different values of $\beta$k along with a=0.7, b=1.2, d=2, M=2, $\phi$=$\frac{\pi}{4}$  \\
\includegraphics[width=4.8in,height=3.2in]{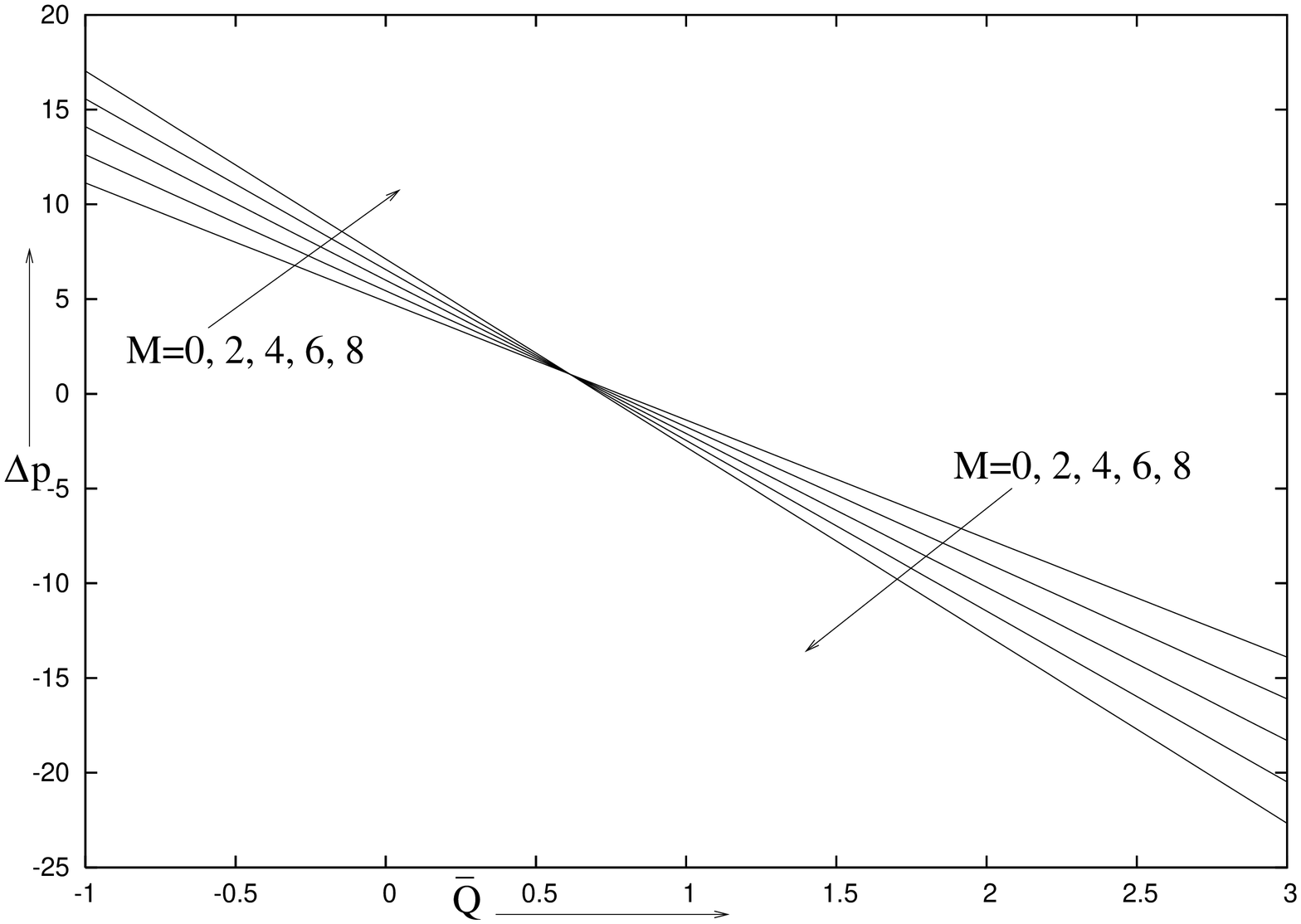} \\
 Fig. 8: Variation of $\bar{Q}$ with  $\Delta p$ for different values of M along with a=0.7, b=1.2, d=2, $\beta$k=0.1, $\phi$=$\frac{\pi}{4}$ \\
\end{center}
\end{minipage}\vspace*{.25cm}

\newpage

\begin{minipage}{1.0\textwidth}
     \begin{center}
 \includegraphics[width=5.0in,height=3.8in]{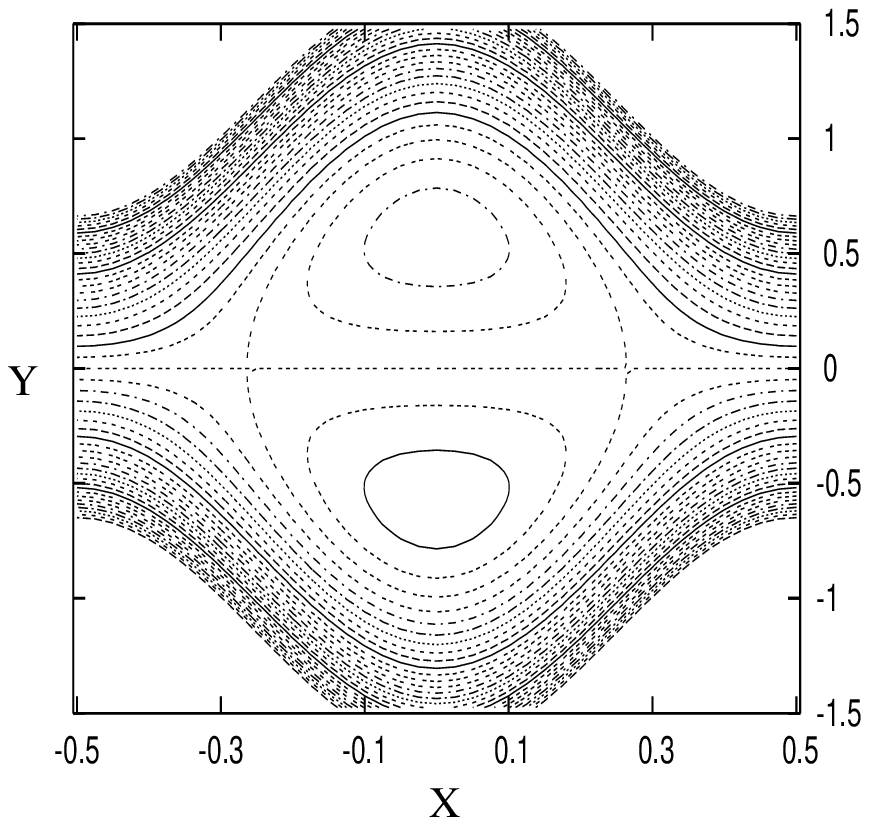}\\
                   (a)\\
\includegraphics[width=5.0in,height=3.8in]{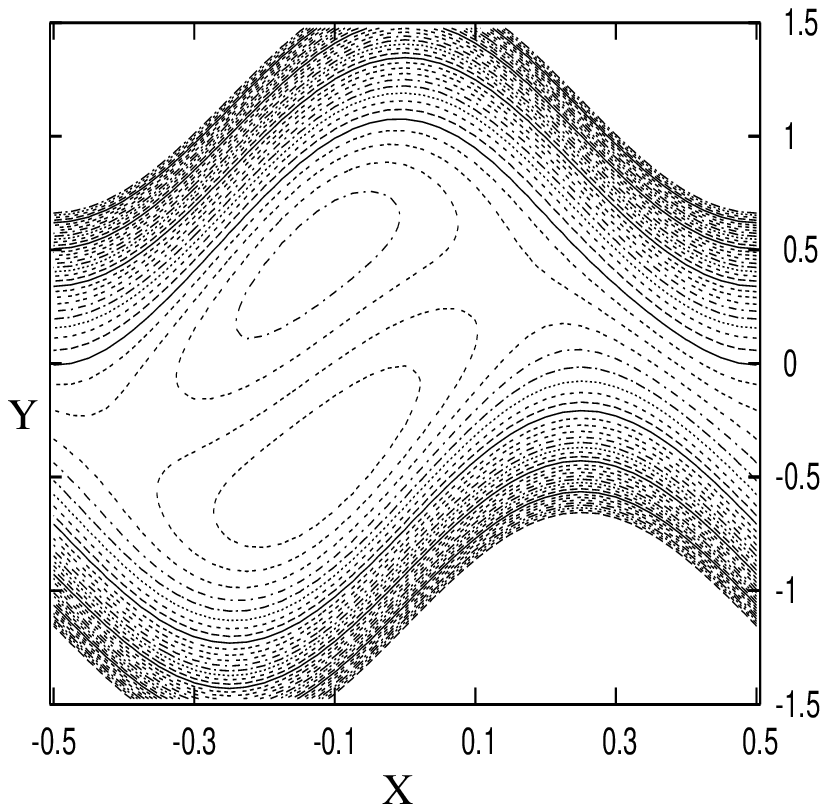} \\
         (b)\\
\end{center}
\end{minipage}\vspace*{.25cm}
\newpage
\begin{minipage}{1.0\textwidth}
     \begin{center}
 \includegraphics[width=5.0in,height=3.8in]{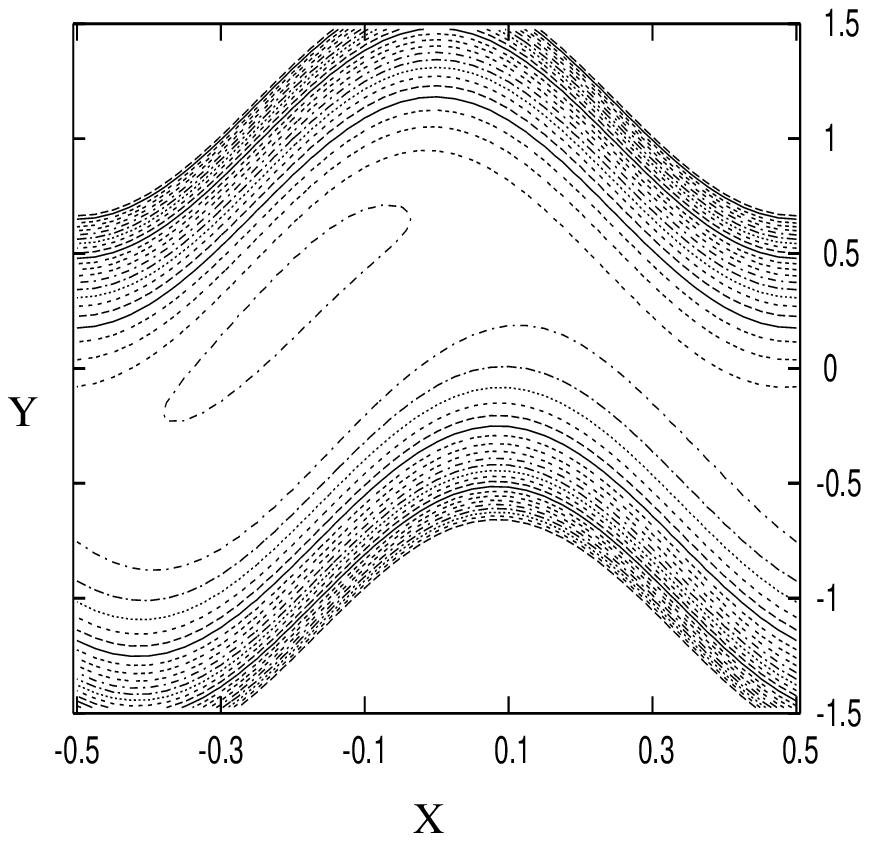}\\
                (c)\\
 \includegraphics[width=5.0in,height=3.8in]{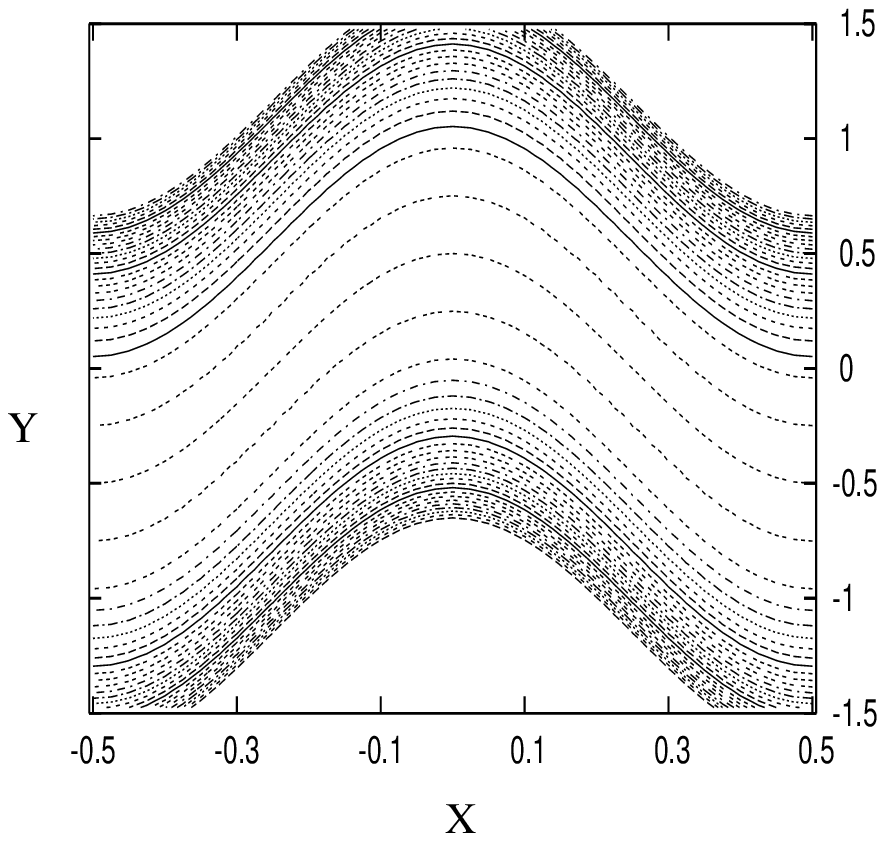} \\
  (d)\\ 

Figs. 9: Streamlines when a=b=0.5, $\beta k=1.0$, d=1, M=1 (a) $\phi$=0.0, (b) $\phi$=$\frac{\pi}{2}$,\\ (c) $\phi$=$\frac{5\pi}{6}$, (d) $\phi$=$\pi$
\end{center}
\end{minipage}\vspace*{.25cm}
\newpage

\begin{minipage}{1.0\textwidth}
     \begin{center}
 \includegraphics[width=5.0in,height=3.8in]{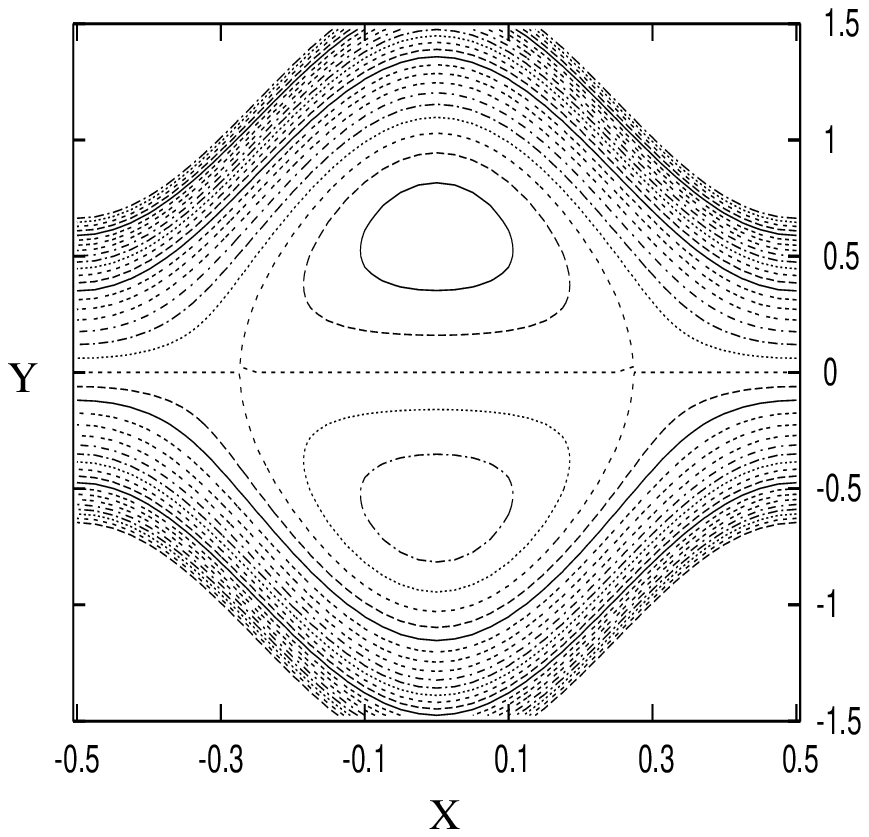}\\
     (a)\\
\includegraphics[width=5.0in,height=3.8in]{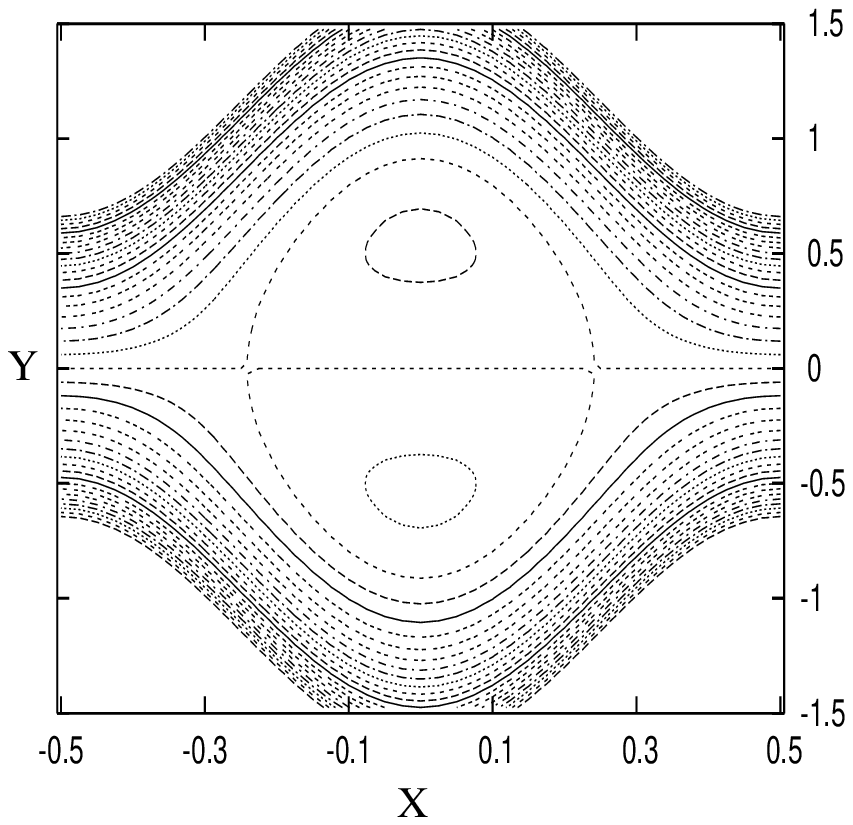} \\
       (b)\\ 
\end{center}
\end{minipage}\vspace*{.25cm}
\newpage

\begin{minipage}{1.0\textwidth}
     \begin{center}
 \includegraphics[width=5.0in,height=3.8in]{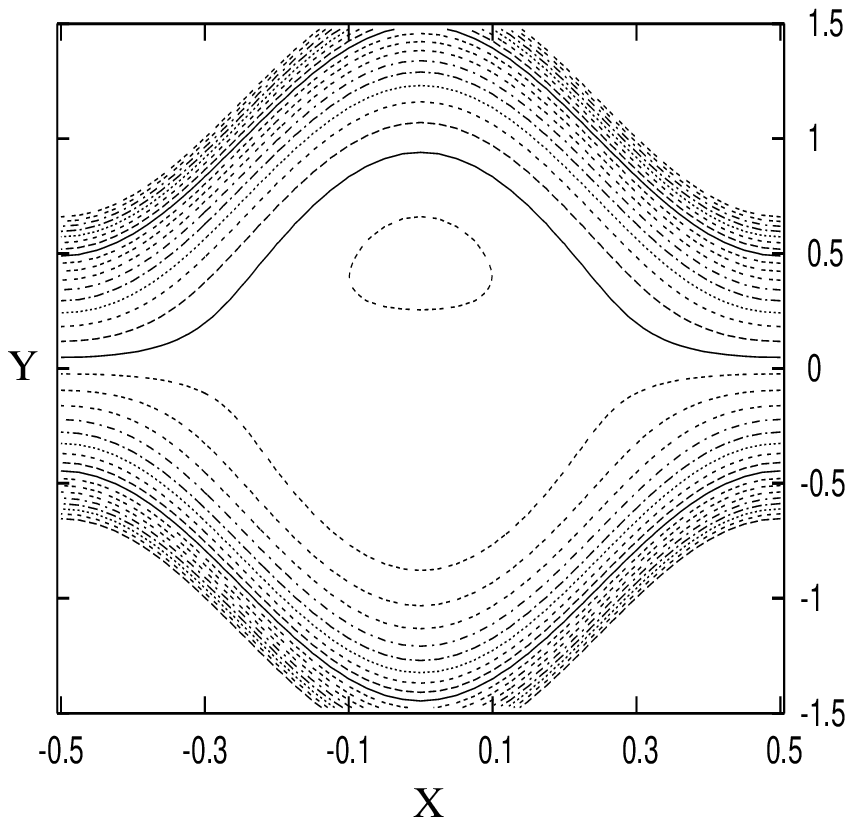}\\
                (c) \\
 \includegraphics[width=5.0in,height=3.8in]{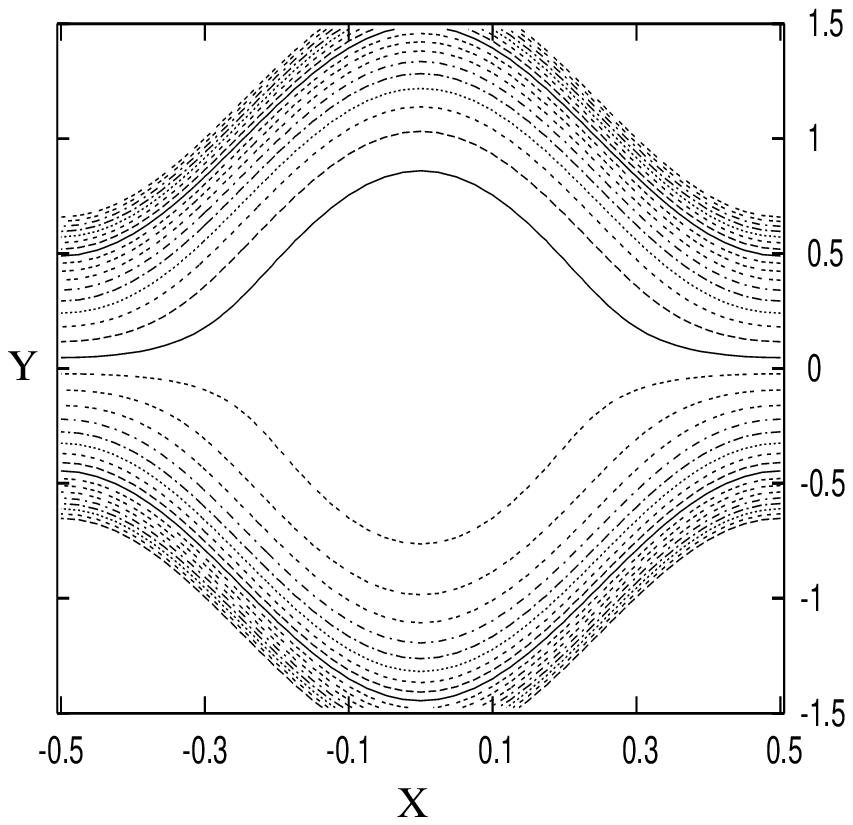} \\
                       (d)\\ 
Figs. 10: Streamlines when a=b=0.5, $\beta k=1.0$, $\phi =0.0$, d=1.0  (a) M=0.0, (b) M=3.0,\\ (c) M=5.0, (d) M=7.0
\end{center}
\end{minipage}\vspace*{.25cm}
\newpage

\begin{minipage}{1.0\textwidth}
     \begin{center}
 \includegraphics[width=5.0in,height=3.8in]{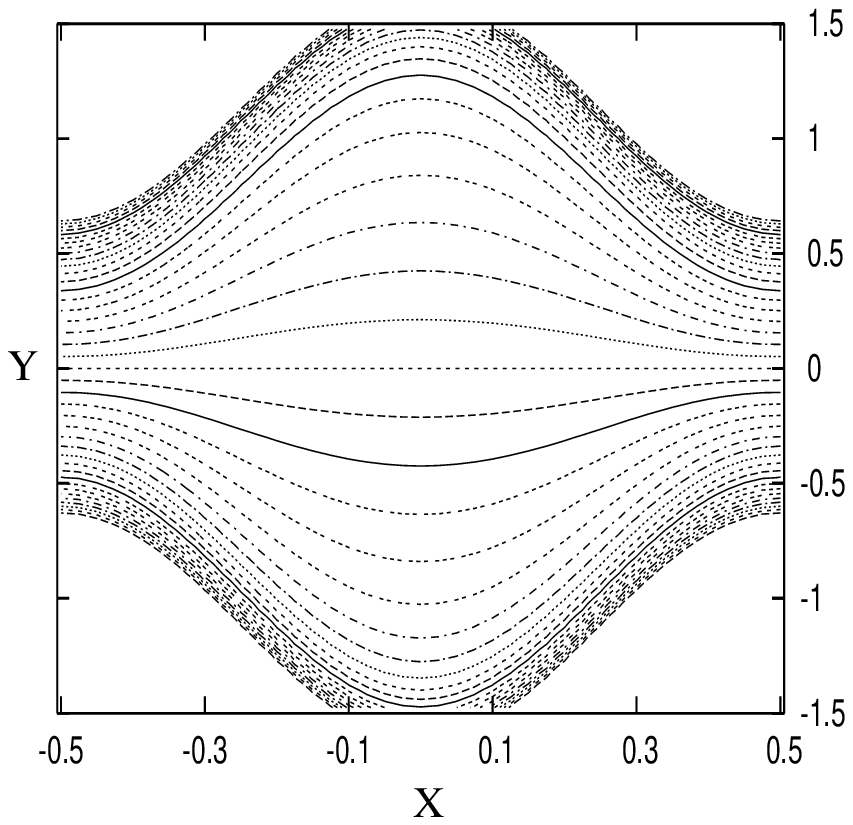}\\
                  (a)\\
\includegraphics[width=5.0in,height=3.8in]{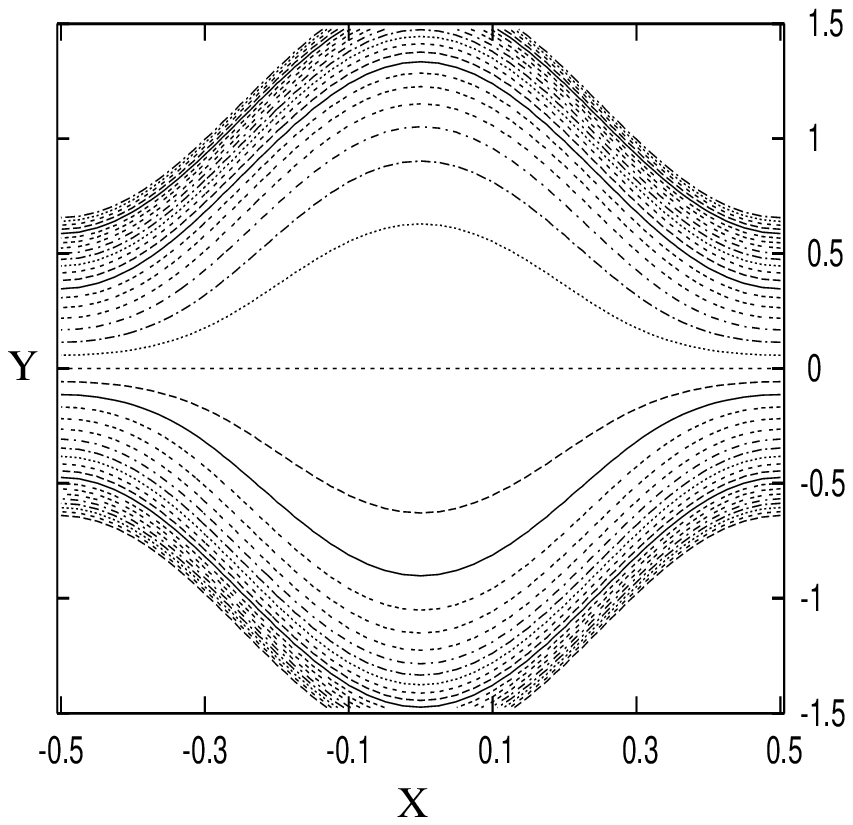} \\
         (b)\\ 
\end{center}
\end{minipage}\vspace*{.25cm}
\newpage

\begin{minipage}{1.0\textwidth}
     \begin{center}
 \includegraphics[width=5.0in,height=3.8in]{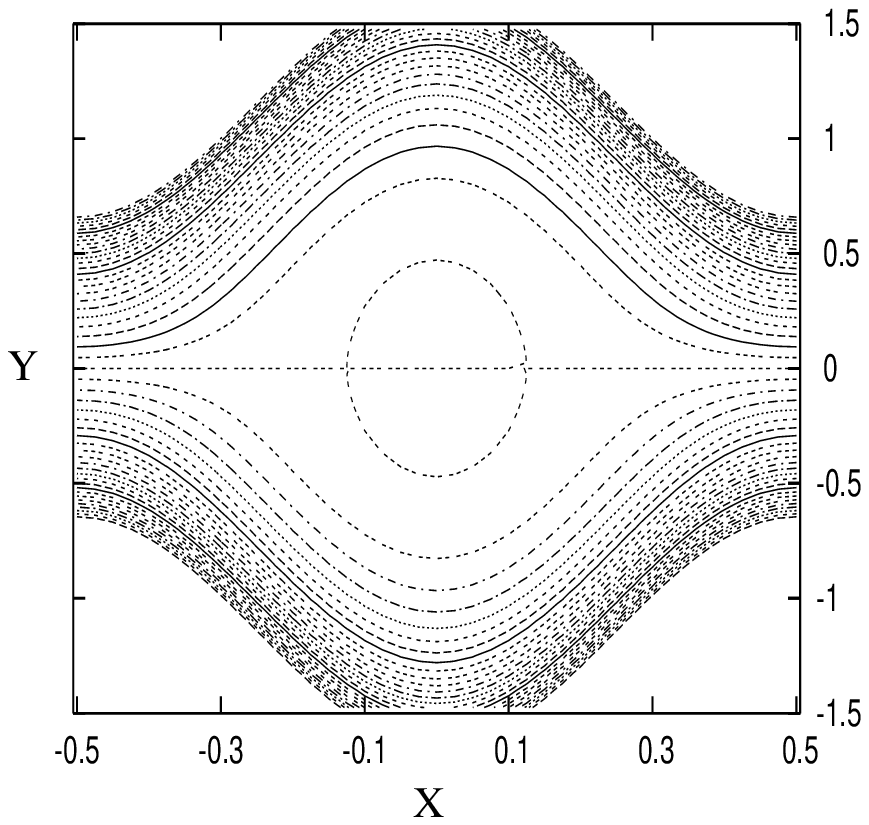}\\
             (c) \\
 \includegraphics[width=5.0in,height=3.8in]{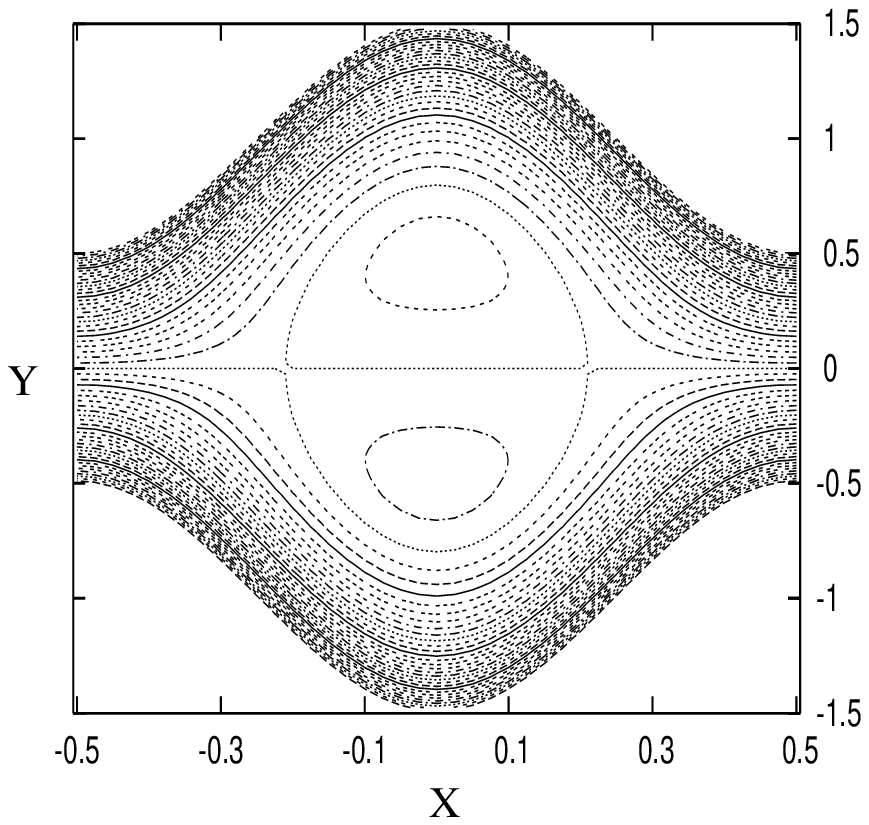} \\
                       (d)\\ 
Figs. 11: Stream lines when a=b=0.5, M=1.0, $\phi =0.0$, d=1  (a) $\beta$k=0.02, (b) $\beta$k=0.08,\\ (c) $\beta$k=0.13, (d) $\beta$k=0.15
\end{center}
\end{minipage}\vspace*{.25cm}
\newpage

\begin{minipage}{1.0\textwidth}
     \begin{center}
 \includegraphics[width=5.0in,height=3.8in]{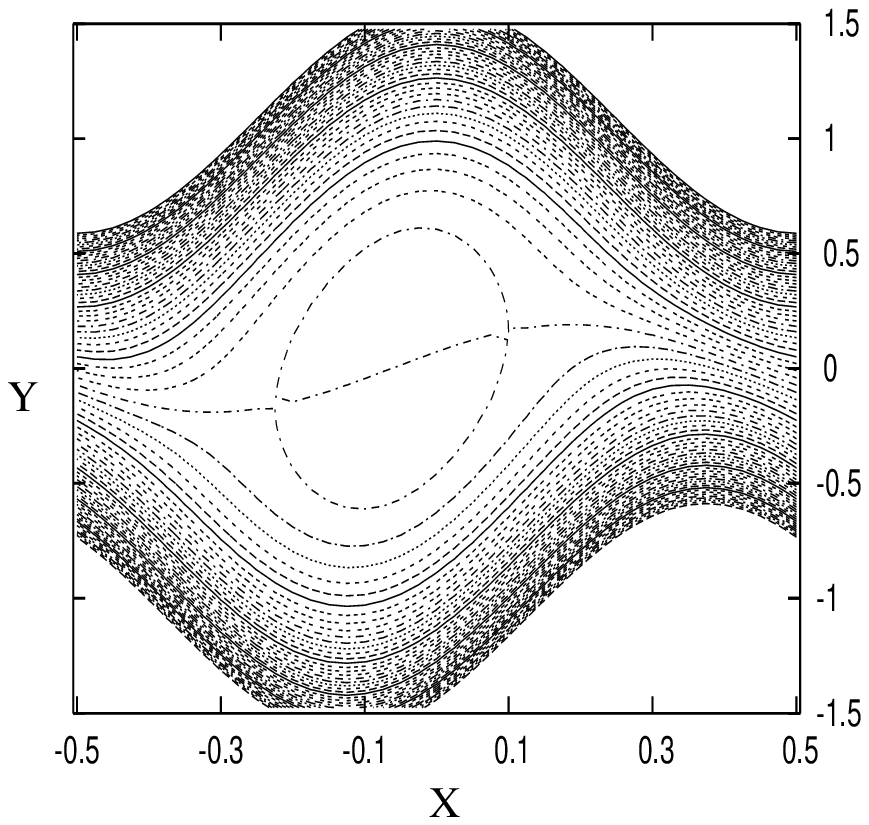}\\
          (a) \\
\includegraphics[width=5.0in,height=3.8in]{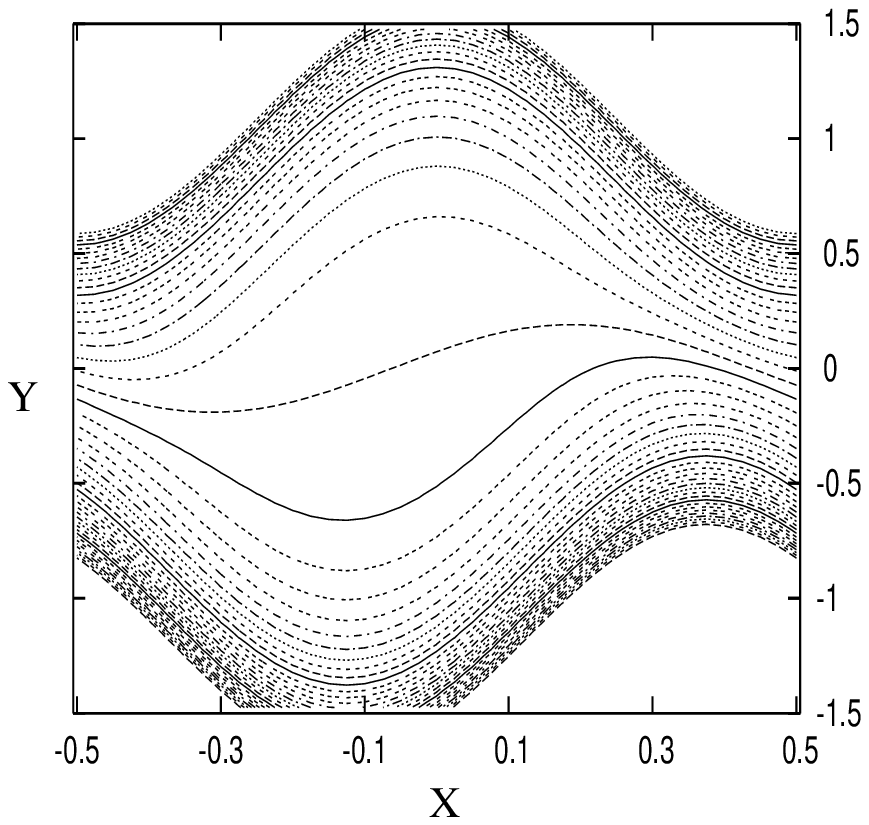} \\
               (b)\\ 
Figs.: 12: Stream lines when a=b=0.5, $\beta$k=0.15, $\phi$=$\frac{\pi}{4}$, d=1  (a) M=1, (b) M=5
\end{center}
\end{minipage}\vspace*{.25cm}

\newpage
\begin{minipage}{1.0\textwidth}
     \begin{center}
 \includegraphics[width=5.0in,height=3.8in]{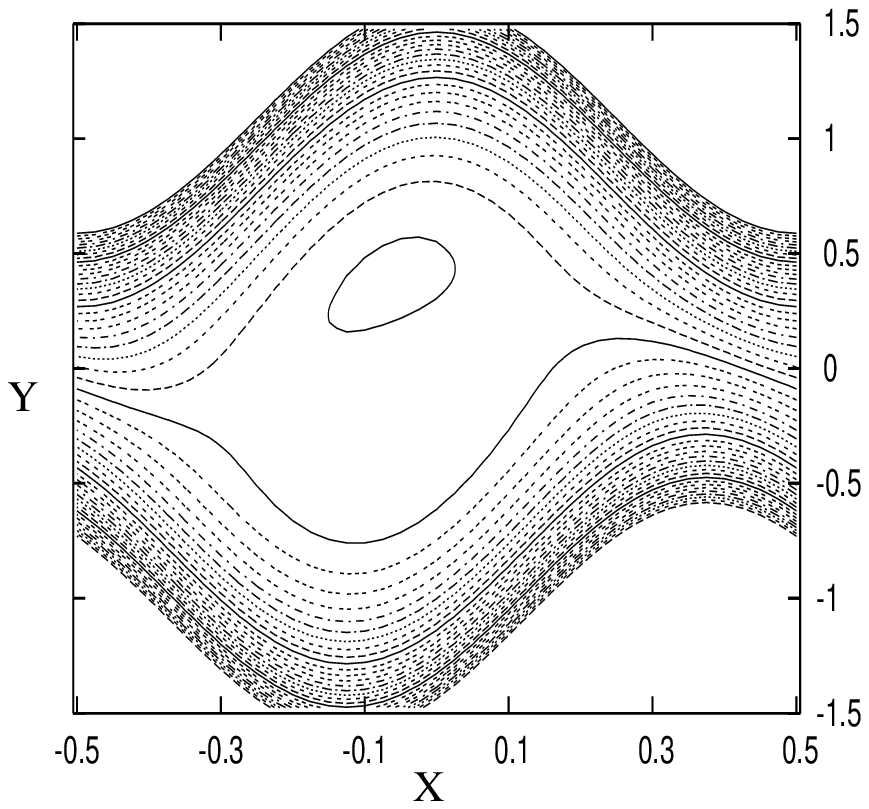}\\
       (a)\\
\includegraphics[width=5.0in,height=3.8in]{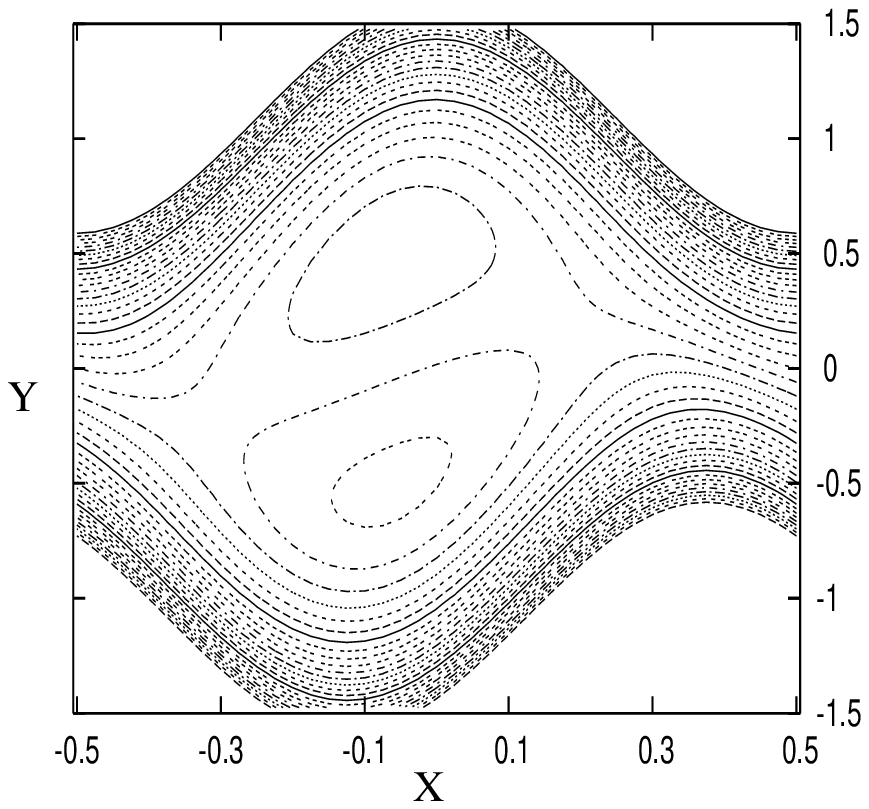} \\
          (b)\\ 
Figs.: 13: Stream lines when a=b=0.5, M=2, $\phi$=$\frac{\pi}{4}$, d=1  (a) $\beta$k=0.2, (b) $\beta$k=0.5
\end{center}
\end{minipage}\vspace*{.25cm}

\newpage
\begin{minipage}{1.0\textwidth}
     \begin{center}
 \includegraphics[width=5.0in,height=3.8in]{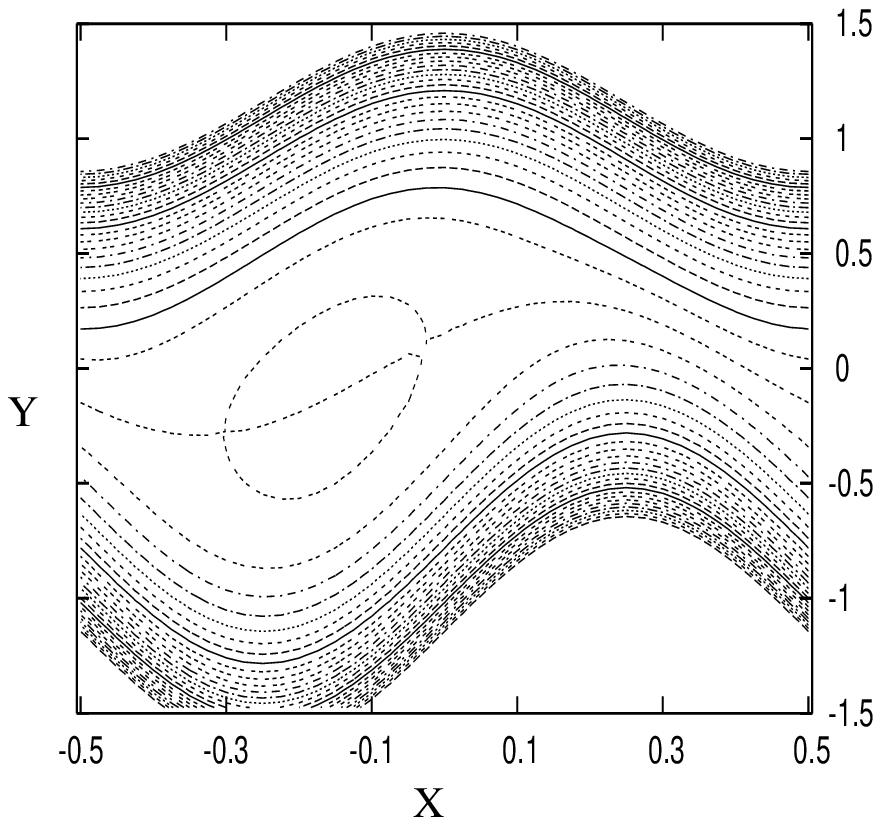}\\
             (a) \\
\includegraphics[width=5.0in,height=3.8in]{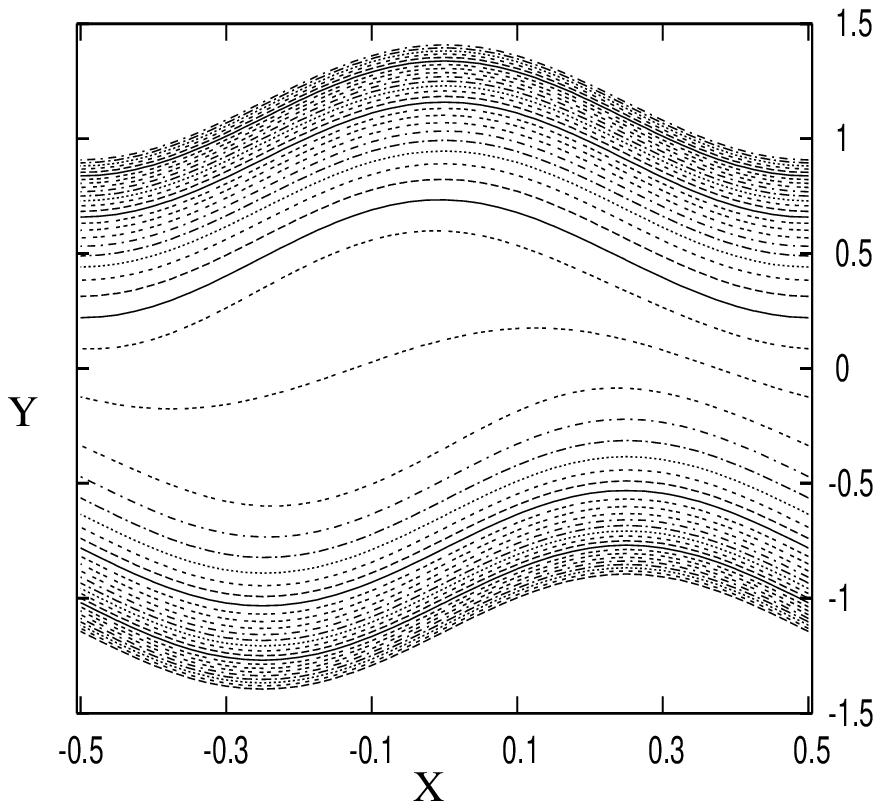} \\
             (b)\\ 
\end{center}
\end{minipage}\vspace*{.25cm}
\newpage
\begin{minipage}{1.0\textwidth}
     \begin{center}
 \includegraphics[width=5.0in,height=3.8in]{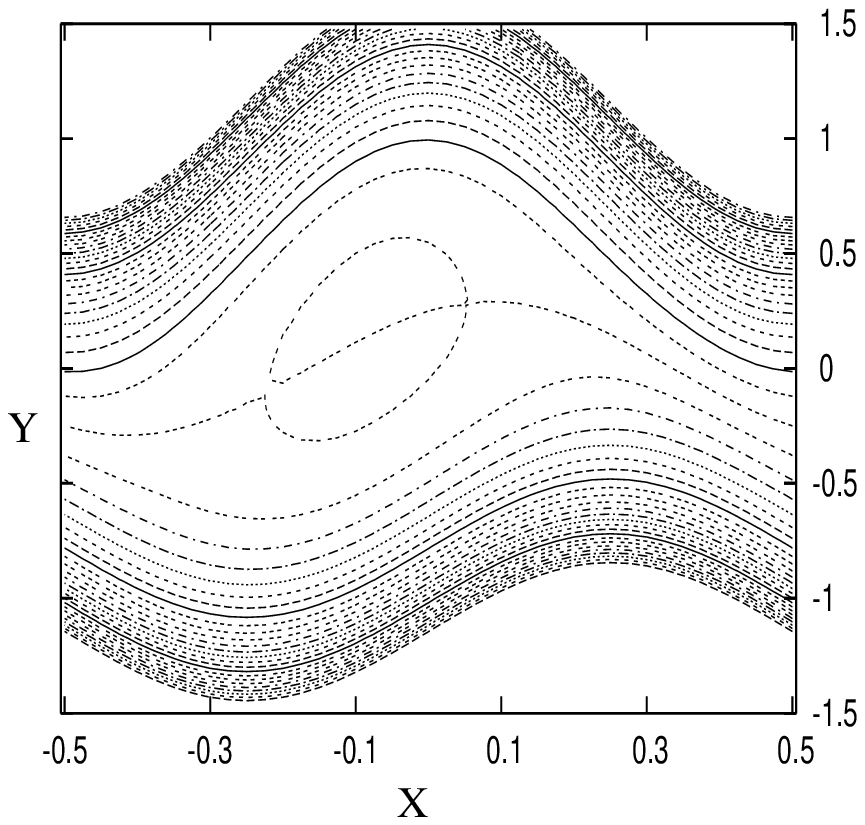}\\
               (c) \\ 
Figs. 14: Stream lines when M=2, $\beta$k=0.2, $\phi$=$\frac{\pi}{2}$, d=1  (a) a=0.3, b=0.5 (b) a=0.4, b=0.4\\ (c) a=0.5, b=0.3
\end{center}
\end{minipage}\vspace*{.25cm}

\end{document}